\documentclass[twocolumn]{article}
\usepackage{amsmath,amssymb,amsfonts}
\usepackage[version=4]{mhchem}
\usepackage{siunitx}
\usepackage[sort&compress,super,square,comma]{natbib}
\usepackage{graphicx}
\usepackage{booktabs}
\usepackage{subcaption}
\usepackage{xcolor}

\usepackage[a4paper,left=1.5cm,right=1.5cm,top=3cm,bottom=4cm]{geometry}
\usepackage[font=footnotesize]{caption}
\makeatletter
\newcommand{\@defs@vec}[1]{\boldsymbol{#1}}
\newcommand{\@defs@tens}[1]{\mathbf{#1}}

\newcommand{\tens}[1]{\@defs@tens{#1}{}}      

\DeclareMathOperator{\tr}{tr}        

\newcommand{\trp}{\textsf{T}}        

\newcommand{\divgO}{\nabla_0\!\cdot\!}   

\newcommand{\Li}{\text{Li}}      
\newcommand{\LiO}{\text{Li,0}}      
\newcommand{\rev}{\text{rev}}      
\newcommand{\rf}{\text{ref}}      
\newcommand{\pla}{\text{pl}}      

\newcommand{\SEI}{\text{SEI}}
\newcommand{\ely}{\text{elyt}}
\newcommand{\Lint}{{\text{Li}^0}}
\newcommand{\LintO}{{\text{Li}^0,0}}
\newcommand{\dev}{\text{dev}}

\newcommand{\inner}{\text{in}}
\newcommand{\out}{\text{out}}

\newcommand{\cref}{c_\text{max}}

\newcommand{\ch}{\text{ch}}
\newcommand{\el}{\text{el}}

\makeatother

\begin{document}

\title{Chemo-Mechanical Model of SEI Growth on Silicon Electrode Particles}

\date{\today}

\author{Lars von Kolzenberg \footnotemark[2] \footnotemark[3]
\and Arnulf Latz \footnotemark[2] \footnotemark[3] \footnotemark[4]
\and Birger Horstmann  \thanks{Corresponding Author: birger.horstmann@dlr.de}
\thanks{German Aerospace Center, Pfaffenwaldring 38-40, 70569 Stuttgart, Germany} \thanks{Helmholtz Institute Ulm, Helmholtzstra{\ss}e 11, 89081 Ulm, Germany} \thanks{Ulm University, Albert-Einstein-Allee 47, 89081 Ulm, Germany} }

\maketitle

\begin{abstract}

Silicon anodes promise high energy densities of next-generation lithium-ion batteries, but suffer from shorter cycle life.
The accelerated capacity fade stems from the repeated fracture and healing of the solid-electrolyte interphase (SEI) on the silicon surface.
This interplay of chemical and mechanical effects in SEI on silicon electrodes causes a complex aging behavior.
However, so far, no model mechanistically captures the interrelation between mechanical SEI deterioration and accelerated SEI growth.
In this article, we present a thermodynamically consistent continuum model of an electrode particle surrounded by an SEI layer. 
The silicon particle model consistently couples chemical reactions, physical transport, and elastic deformation. 
The SEI model comprises elastic and plastic deformation, fracture, and growth. 
Capacity fade measurements on graphite anodesand in-situ mechanical SEI measurements on lithium thin films provide parametrization for our model.
For the first time, we model the influence of cycling rate on the long-term mechanical SEI deterioration and regrowth. 
Our model predicts the experimentally observed transition in time dependence from square-root-of-time growth during battery storage to linear-in-time growth during continued cycling.
Thereby our model unravels the mechanistic dependence of battery aging on operating conditions and supports the efforts to prolong the battery life of next-generation lithium-ion batteries. \\
\end{abstract}

\section{Introduction}

Lithium-ion batteries progressed to the benchmark battery technology for mobile applications owing to their superior energy density as well as longevity. 
The use of silicon anodes would further increase the energy density, because silicon has nearly the tenfold theoretical capacity of the currently used graphite \cite{Chan2008,Sun2016}. However, this capacity gain comes at the cost of volume expansions up to 300\% \cite{Zhang2011}. These large expansions lead to high mechanical stresses, which deteriorate the anode and lead to faster aging and shorter battery lifetime  \cite{Zhang2011,Wu2012a}.

The main aging mechanism in lithium-ion batteries with graphite or silicon anodes is the formation and growth of the solid-electrolyte interphase (SEI) \cite{Peled1979,Peled1995,Peled1997,Horowitz2018a,Li2018,Delpuech2013,Jerliu2018,Mazouzi2012,Steinruck2020,li2018single,Kalaga2018}. The SEI forms during the first battery cycle, when the anode potential is drawn below the electrolyte reduction potential \cite{Goodenough2010,Horstmann2018,Wang2018}.
This initiates reactions of electrolyte molecules with lithium ions, which form organic compounds like lithium ethylene dicarbonate $\ce{Li_2EDC}$ and inorganic compounds like $\ce{LiF}, \ce{Li_2CO_3}$, and $\ce{Li_2O}$ \cite{lu2014chemistry,Peled1997,Aurbach1999,Winter2009,Xu2004,Agubra2014,An2016,Lu2011,Xu2014,Wang2018,Huang2019c,Boniface2016a,Tokranov2014,Tokranov2016,Kumar2016,Veith2017}. These products precipitate on the anode in a dual layer structure with a compact, inorganic inner layer and a porous, organic outer layer \cite{Edstrom2006,Peled2017} and thus form a nanometer thick and stable SEI at around $\SI{0.15}{\volt}$ vs. lithium metal \cite{Veith2017}.
In subsequent cycles, this SEI shields the electrolyte from the low anode potentials and thereby enables a stable battery operation. 
However, the shielding effect is not perfect, so that the SEI continues to grow over time effectively lowering the usable capacity \cite{Peled1995,Peled1997}.

Battery storage experiments revealed that long-term SEI growth follows a $\sqrt{t}$-time dependence pointing to a self-passivating process \cite{Broussely2001}. 
As a possible long-term growth mechanism electrolyte diffusion \cite{Single2018,Ploehn2004,Single2017,Single2016,Pinson2012,Tang2011,Tang2012a,Tang2012b,Tahmasbi2017,Hao2017,Roder2016,Tokranov2016}, electron conduction \cite{Single2018,VonKolzenberg2020,Staniewicz2005,Single2017,Pinson2012,Tang2012b,Christensen2004,Colclasure2011,Roder2017,Das2019}, electron tunneling \cite{Single2018,VonKolzenberg2020,Li2015} and the diffusion of neutral lithium interstitial atoms \cite{Single2018,VonKolzenberg2020,Shi2013,Soto2015} were proposed. However, only the diffusion of electrons, \textit{e.g.}, via neutral lithium interstitial atoms, yields the experimentally observed voltage dependence of capacity fade \cite{Keil2016,Single2018}.

Besides the open circuit voltage of the anode, also the operating conditions during battery cycling strongly affect SEI growth. In a recent experiment, Attia \textit{et al.} \cite{Attia2019} showed the dependence of SEI growth on the magnitude and direction of applied current. Two mechanistic models describe this experimentally observed trend with good accuracy \cite{VonKolzenberg2020,Das2019}.
Implementing mechanistic SEI models to three-dimensional (3D) cell simulations gives further insights on the effect of battery operation on SEI growth.
Using a purely reaction limited model for SEI growth Franco and coworkers \cite{Chouchane2021} evaluated SEI heterogeneity in a 3D model, and discussed its dependence on electrode mesostructure. By leaving out the coupling to transport limitations this approach is only valid for the first few cycles contrary to the fully coupled models of Bazant, Horstmann, and Latz \cite{Das2019,Schmitt2019,9049261,VonKolzenberg2020}.

Moreover, cycling experiments regularly reveal accelerated SEI growth on high capacity anodes like silicon due to large particle expansion and contraction (breathing) \cite{Peled2017,McBrayer2021}. This geometrical change strains the SEI until it eventually fractures, which leads to formation of new SEI upon direct contact between electrolyte and electrode~\cite{Kumar2016,Kumar2017,Jangid2017,Guo2020}. Several groups developed mechanistic models to describe the mechanical response of the SEI on battery cycling \cite{Deshpande2012,Xu2015,Deshpande2017a,Rejovitzky2015b,Kamikawa2021,Guo2020,li2018single,Verma2017,Galvez-Aranda2019c,Verbrugge2015,He2015,He2017,Laresgoiti2015,Deng2019,Zhao2012,Tanaka2018,Kotak2018}. 
However, these models focus on SEI mechanics and incorporate at most simple SEI growth models \cite{Deshpande2012,Xu2015,Deshpande2017a,Rejovitzky2015b,li2018single,Laresgoiti2015,Kotak2018}.

In this paper, we develop a detailed electrochemo-mechanical model to describe SEI mechanics and growth on a deforming electrode particle. We describe the chemomechanics of the electrode particle with a thermodynamically consistent model \cite{Castelli2021}. The electrochemical part of the SEI model relies on our previous works on SEI growth \cite{Single2016,Single2017,Single2018,VonKolzenberg2020}. The mechanical part of the SEI model describes the SEI as porous dual-layer structure \cite{Edstrom2006,Peled2017}, which deforms elastic-perfectly plastic \cite{Yoon2020a}.

In the next section, we develop the model based on irreversible thermodynamics and show the details of its implementation in the subsequent section. Afterwards, we parametrize the model chemistry and mechanics with recent experiments \cite{Keil2016,Yoon2020a}. Based on this parametrization we analyze the electrochemical and mechanical predictions of the model in the short- and long-term. Finally, we summarize our results and show possible extensions of our model.

\section{Theory}
This section briefly presents our model for coupling chemistry and mechanics in electrode particles covered by SEI.
The reader is referred to section SI-1 in the supporting information for further details. We summarize the system of differential-algebraic equation, which we implement, in Equation SI-45 to Equation SI-51.

\begin{figure}[tb] 
 \centering
 \includegraphics[width=8.4cm]{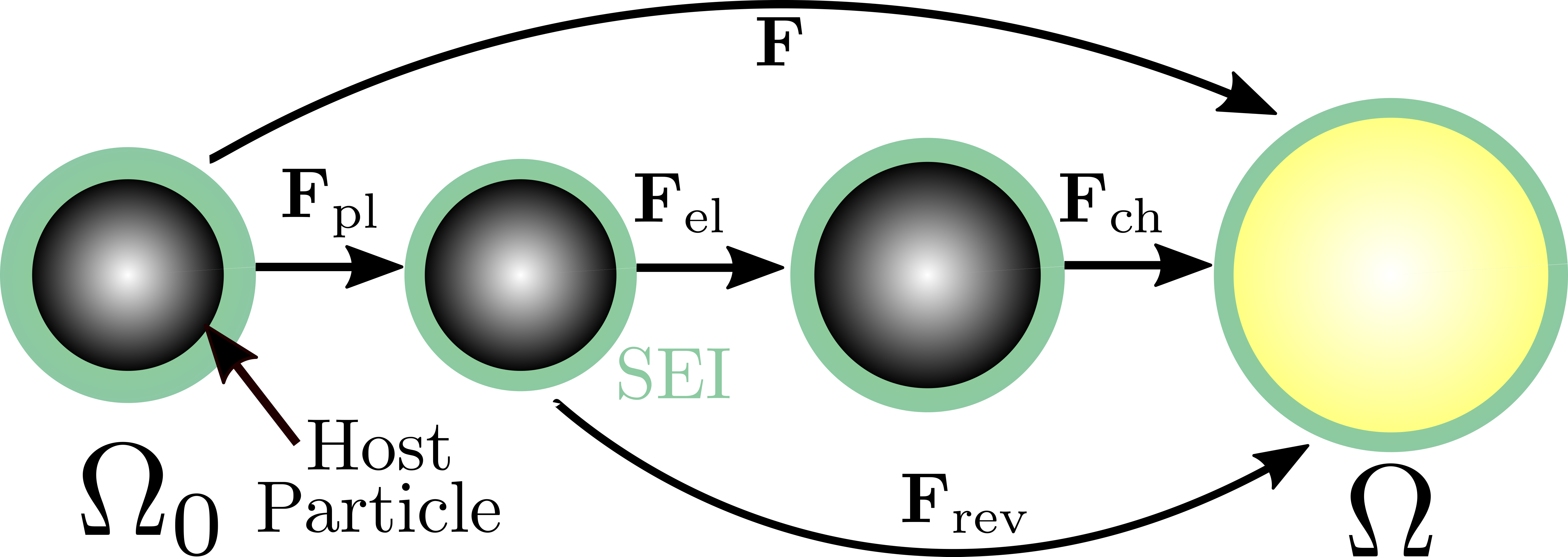}
 \caption{Schematic representation of particle and SEI deformation. The compound deforms plastically $\tens{F}_\pla$, elastically $\tens{F}_\el$, and finally chemically $\tens{F}_\ch$.}
 \label{fig:Schematic_Deformation}
\end{figure}

The deformations of the silicon electrode particle and the SEI during lithium intercalation and deintercalation is schematically depicted in Figure \ref{fig:Schematic_Deformation}. 
These are captured with deformation gradient $\tens{F}=\partial \vec{x}/\partial \vec{X}_0$, which relates the Lagrangian domain $\Omega_0$ to the Eulerian domain $\Omega$. The volume expansion is given by its determinant $\det \tens{F} = J=V/V_0$. We divide the deformation into a reversible $\tens{F}_\rev$ and an irreversible part $\tens{F}_\pla$ \cite{holzapfel2000nonlinear}.

We derive our model from non-equilibrium thermodynamics in Section SI-1 and find the following inequality for the dissipation $\mathcal{R}_0$ in the Lagrangian frame, \cite{Latz2011,Latz2015,Schammer2021,Castelli2021,Single2019}
\begin{align}
    \label{eq:Dissipation_Final}
    \begin{split}
        \mathcal{R}_0 &=  - \vec{N}_\LiO \cdot \divgO \mu_\Li +\tens{M}\colon\tens{L}_\pla \geq 0.
    \end{split}
\end{align}
Equation \ref{eq:Dissipation_Final} restricts the choice of constitutive equations for lithium flux $\vec{N}_\LiO$ and plastic flow $\tens{L}_\pla=\dot{\tens{F}}_\pla \tens{F}_\pla^{-1}$. They depend on the corresponding forces, namely the chemical potential $\mu_\Li$ and the Mandel stress $\tens{M}=J \tens{F}_\rev^\trp \sigma \tens{F}_\rev^{-\trp}$ with the Cauchy stress $\sigma$ \cite{mandel1972plasticite,Lubliner2006}. In the following, we present our electrode particle and SEI growth models.

\subsection{Electrode Particle Model}
We use the chemo-mechanical electrode particle model discussed in our previous work \cite{Castelli2021}. The particle deformation $\tens{F}$ is completely reversible and consists of an elastic part $\tens{F}_\el$ due to mechanical stress and a chemical part $\tens{F}_\ch$ resulting from changes in lithium concentration $c_\LiO$,
\begin{equation}
\label{eq:Deformation_Particle}
\tens{F}=\tens{F}_\rev= \tens{F}_\el\tens{F}_\ch.
\end{equation}
The continuity equation \ref{eq:Continuity_Equation_Particle}, 
\begin{equation}
 \label{eq:Continuity_Equation_Particle}
    \dot{c}_\LiO= -\divgO \vec{N}_\LiO \quad\text{with} \quad\vec{N}_\LiO = - D_\Li \tfrac{\partial \mu_\Li}{\partial c_\LiO} \nabla_0 \mu_\Li,    
\end{equation}
defines the change of lithium concentration with the diffusion constant $D_\Li$. The elastic deformation is determined by the momentum balance 
\begin{equation}
    \label{eq:Momentum_Balance_Particle}
    0 = \divgO \tens{P},    
\end{equation}
with the first Piola--Kirchhoff stress tensor $\tens{P}=J\sigma\tens{F}^{-\trp}$.

\subsection{SEI Model}
Now, we derive a model for coupled SEI growth and mechanics.
We model the SEI as porous medium consisting of an incompressible SEI matrix and electrolyte inside its pores, as introduced by Single \textit{et al.}
\cite{Single2016,Single2017}.
The volume fraction of solid SEI, $\epsilon_\SEI = V_\SEI / V$, characterizes the morphology at each point.

The deformation tensor $\bar{\tens{F}}$ describes the  overall volume deformation and consists of three parts,
\begin{equation}
\label{eq:Deformation_SEI}
\bar{\tens{F}} =   \bar{\tens{F}}_\el\bar{\tens{F}}_\pla\bar{\tens{F}}_\rf.
\end{equation}
The first part is a reference deformation $\bar{\tens{F}}_\rf$, which we introduce to set the stress free SEI configuration. Most importantly, the SEI deforms plastically $\bar{\tens{F}}_\pla$ and elastically $\bar{\tens{F}}_\el$.

\subsubsection*{SEI Growth}
\label{sec:SEIGrowth}
The SEI grows when electrons $\ce{e^-}$, lithium ions $\ce{Li^+}$, and electrolyte molecules react. 
We simplify the multitude of possible SEI growth reactions to the formation of the most prominent SEI component $\ce{Li_2EDC}$ from $\ce{EC}$,
\begin{equation}
\label{eq:SEI_reaction}
2\ce{Li^0}+2\ce{EC} \rightarrow \ce{Li_2EDC} +  \ce{R}
\end{equation}
with neutral lithium interstitials $\ce{Li^0}=\ce{Li^+}+\ce{e^-}$, the solvent EC, and gaseous byproducts $\ce{R}$.

We model the transport of electrons through the SEI as diffusion of localized electrons according to our previous works \cite{Single2018,VonKolzenberg2020}. A prominent example are the aforementioned lithium interstitials  $\ce{Li^0}$ \cite{Shi2013,Soto2015}. 
The \ce{Li^0} concentration in the SEI evolves according to
\begin{equation}
\label{eq:Li_Concentration_Differential}
\frac{\text{d}(\epsilon_\SEI c_\LintO)}{\text{d}t}=- \epsilon_\SEI \divgO  \vec{N}_\LintO- 2 \bar{J} r_\SEI \Gamma A_\text{V}.
\end{equation}
Here, the diffusion of \ce{Li^0} through the SEI is described by the flux density $\vec{N}_\Lint$, see subsection SI-1.3. The diffusing \ce{Li^0} can react with the solvent inside the SEI pores and at the SEI-electrolyte interface. We model the available reaction surfaces with the specific surface area $A_\text{V}$ adapted from Single \textit{et al.} \cite{Single2016,Single2017} and shown detailedly in section SI-4. In Equation \ref{eq:Li_Concentration_Differential}, $\Gamma$ is the surface site density and $r_\SEI$ is the rate of the SEI formation reaction in Equation \ref{eq:SEI_reaction},
\begin{equation}
    \label{eq:Reaction_Rate}
    r_\SEI = k c_\Lint^2.
\end{equation}
Here, $c_\Lint$ is the \ce{Li^0} concentration inside the SEI and $k$ is a rate constant.

We capture formation of new SEI by tracking the SEI volume fraction in the Lagrangian frame $\varepsilon_{\SEI,0}=\bar{J}\epsilon_\SEI$,
\begin{equation}
\label{eq:Porosity_Change_Differential}
\dot{\varepsilon}_{\SEI,0} = \bar{J} r_\SEI \Gamma A_\text{V}\bar{V}_\SEI
\end{equation}
with the average molar volume of SEI components $\bar{V}_\SEI$. In summary, our SEI growth model takes into account the coupling of transport and reaction processes.

\subsubsection*{SEI Mechanics}\label{sec:SEIMechanics}
The SEI deforms elastoplastically until it eventually fractures as the electrode particle beneath expands and contracts. We determine the elastic deformation with the momentum balance inside the SEI
\begin{equation}
    \label{eq:Momentum_SEI}
    \divgO \tens{P}_\SEI = 0.
\end{equation} 
and determine $\tens{P}_\SEI$ from the micromechanical model for porous solids of Danielsson \textit{et al.} \cite{Danielsson2004}, see Equation SI-37.

Based on the stress $\tens{P}_\SEI$, we proceed to develop a model for plastic deformation and fracture of the SEI. For the plastic deformation we introduce the yield function $f$, which tends to zero if the SEI reaches its yield criterion. The fracture depends on the damage variable $\xi$, which describes the degree of deterioration and reaches from $0$ (intact) to $1$ (broken). We couple the damage variable to the yield function with the Gurson-Tvergaard-Needleman approach \cite{Gurson1977,Tvergaard1981,Tvergaard1982,Tvergaard1984}
\begin{equation}
\label{eq:yield_function_SEI}
f=\frac{\frac{3}{2}\left|\tens{M}^\dev_\text{c}\right|^2}{\sigma_\text{Y}^2}+2\xi \cosh \left(\frac{1}{2}\frac{\tr\left(\tens{M}_\text{c}\right)}{\sigma_\text{Y}} \right)-1-\xi^2\leq 0.
\end{equation}
Here $\tens{M}_\text{c}^\dev=\tens{M}_\text{c}-1/3\tr\tens{M}_\text{c}$ is the deviatoric part of the adapted Mandel stress $\tens{M}_\text{c}=\tens{F}_\rev^\trp \sigma \tens{F}_\rev^{-\trp}$ inside the SEI and $\sigma_\text{Y}$ is the yield stress \cite{mandel1972plasticite,Lubliner2006,Reusch2003}. The damage variable $\xi$ depends on the SEI porosity $\epsilon_\ely=1-\epsilon_\SEI$ by Equation \ref{eq:Damage_Variable} \cite{Tvergaard1984}.
\begin{equation}
\label{eq:Damage_Variable}
\xi = \begin{cases}  \epsilon_\ely \qquad \text{if $\epsilon_\ely<\epsilon_{\ely,\text{crit}}$}\\
\epsilon_\ely +(1-\epsilon_\ely) \cdot \left(\frac{\epsilon_\ely-\epsilon_{\ely,\text{crit}}}{\epsilon_{\ely,\text{frac}}-\epsilon_{\ely,\text{crit}}}\right) \qquad \text{else}.
\end{cases}
\end{equation}
The critical SEI porosity $\epsilon_{\ely,\text{crit}}$ accounts for pore coalescence, which accelerates mechanical degradation, if the SEI porosity is above the critical porosity $\epsilon_\ely>\epsilon_{\ely,\text{crit}}$. The fracture SEI porosity $\epsilon_{\ely,\text{frac}}$ describes the porosity at which the SEI ultimately breaks with $\xi(\epsilon_\ely\geq\epsilon_{\ely,\text{frac}})=1$.

\begin{figure}[tb] 
 \centering
 \includegraphics[width=8.4cm]{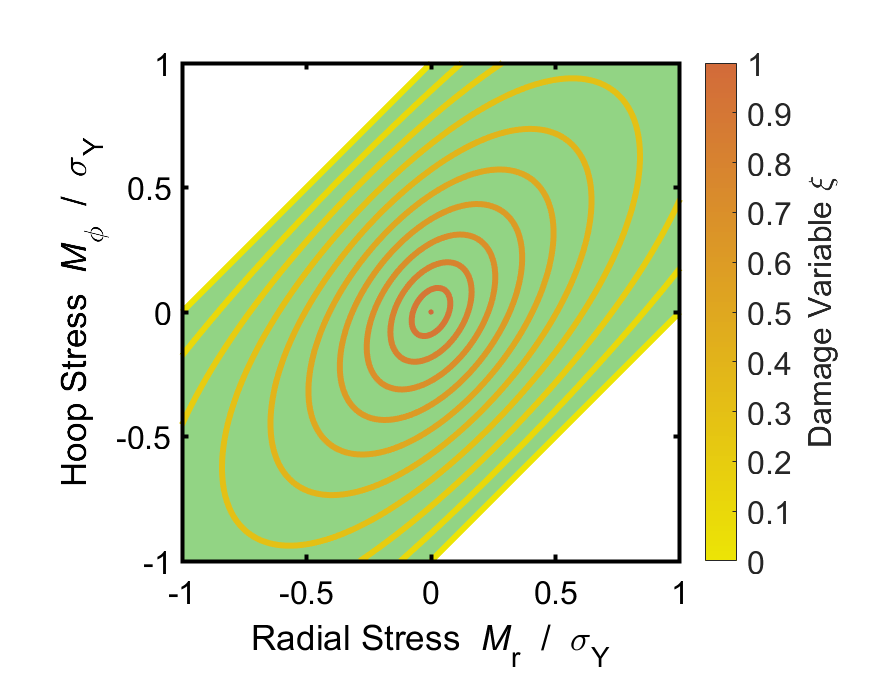}
 \caption{Representation of elastic, plastic and broken regimes in mechanical model. Yield surface $f=0$ for different degrees $\xi$ of damage. The elastic regime for the undamaged case $\xi=0$ is colored in green. With increasing damage $\xi$, the surface shrinks until it converges to $(0,0)$ if SEI is broken $\xi=1$.}
 \label{fig:Yield_Surface}
\end{figure}

In Figure \ref{fig:Yield_Surface}, we show the yield surface $f=0$, at which the SEI flows plastically.
We observe the classical von-Mises yield surface for $\xi=0$, which withstands arbitrary large hydrostatic stress and only depends on the deviatoric stress. Damage causes the yield surface to shrink until it converges to the stress state $(0,0)$ for $\xi=1$.

To describe the plastic flow upon reaching this yield surface, we rely on the maximum plastic dissipation postulate \cite{mises1928mechanik,Hill1947,hill1951state,Taylor1947,mandel1972plasticite,Lubliner2006} as additional restriction to the principle of non-negative dissipation, Equation \ref{eq:Dissipation_Final}. This postulate from plasticity theory constraints plastic flow to the normal direction of the yield surface $\partial f/\partial \tens{M}_\text{c}$ and thus leads to the following constitutive equation \cite{mandel1972plasticite,Lubliner2006},
\begin{equation}
\label{eq:Plastic_flow_SEI}
\tens{L}_\pla = \phi \frac{\partial f}{\partial \tens{M}_\text{c}},
\end{equation}
where the plastic multiplier $\phi$ is non-negative, $\phi \geq 0$, guaranteeing non-negativity of the dissipation rate in Equation \ref{eq:Dissipation_Final}. The plastic multiplier and the yield function $f$ additionally obey the Karush-Kuhn-Tucker condition $\phi f=0$ \cite{Lubliner2006}. Thus, plastic flow is suppressed $\phi=0$ during elastic deformation, $f<0$. For plastic deformation, $f=0$, $\phi$ results from the consistency condition $\dot{f}=0$ \cite{Lubliner2006}. 
Note that the plastic flow is not trace-free and thus not volume preserving, because the yield criterion, Equation \ref{eq:yield_function_SEI} depends on the hydrostatic stress $\tr\left(\tens{M}_\text{c}\right)/3$.

\section{Parameterization}

We assume a homogeneous electrode particle and a dual layer SEI consisting of a dense, inorganic inner layer with a thickness $R_\inner$ and a porous, organic outer layer \cite{Edstrom2006,Peled2017,Single2016,Single2017}. 
We introduce a thickness dependent minimum porosity $\epsilon_{\ely,\text{min}}(\tilde{R})$ to enforce this morphology and stop the SEI reaction once this porosity is reached locally $\tilde{r}_{\SEI,0}(\epsilon_\ely \leq \epsilon_{\ely,\text{min}})=0$. To reduce the set of SEI parameters, we further assume that this minimum porosity corresponds to the critical porosity for pore coalescence $\epsilon_{\ely,\text{crit}}=\epsilon_{\ely,\text{min}}$, see Equation \ref{eq:Damage_Variable}.
Besides the minimum porosity $\epsilon_{\ely,\text{min}}$, also Young's modulus $E_\SEI$ and the yield strength $\sigma_\text{Y}$ vary between both layers.
To continuously transition these SEI properties $y$, we use Equation \ref{eq:Transition},
\begin{align}
\label{eq:Transition}
\begin{split}
&y(R_\inner<R<R_\inner+R_\text{trans})\\
&=y_\inner+(y_\out-y_\inner)\cdot\left(\frac{2 (R-R_\inner)^3}{R_\text{trans}^3}-\frac{3 (R-R_\inner)^2}{R_\text{trans}^2}+1\right)
\end{split}
\end{align}
with the transition thickness $R_\text{trans}$.
In Table SI-1 we list the parameters of our simulation \cite{Single2016,Single2017,Chan2008,Verma2019,Shenoy2010,Single2018,Borodin2006,Naejus1997,Keil2016,Yoon2020a,Obrovac2007}. 
The further parametrization of the SEI on silicon is challenging due to the strong interrelation of electrochemical and mechanical effects. To reduce this complexity, we adopt two experiments, one for each submodel. 
First, we use the experiments of Keil \textit{et al.} \cite{Keil2016} to parametrize the electrochemical part of the SEI model. These experiments were conducted on full cells with a graphite anode during storage so that the influence of mechanics is negligible. Second, the experiments of Yoon \textit{et al.} \cite{Yoon2020a} parametrize the mechanical part of our model. In their experiments, they grow SEI on a thin lithium film and then mechanically strain it. Thus, the experiment is almost independent from electrochemical effects.
The parametrization with other substrates than silicon introduces additional uncertainties to our model. The choice of substrate determines the nucleation properties, but the growth mechanism should be independent from the substrate. However, the evolving SEI composition and structure can, of course, be affected by the innermost nucleation sites of the SEI.

\subsubsection*{Electrochemical SEI Growth}

We rely on the experiments of Keil \textit{et al.} \cite{Keil2016} obtained with graphite anodes to parametrize the chemical SEI growth model. In their experiments, Keil \textit{et al.} \cite{Keil2016} stored batteries at different states of charge and measured the capacity fade after 9.5 months. In line with the approach of Single \textit{et al.} \cite{Single2018}, we subdivide the measured irreversible capacity fade into two parts. The first part $\Delta \text{SoH}_\text{lin}$ is not SEI related and linear in time, the second part stems from the SEI and is predicted by our model.

\begin{figure}[tb]
     \centering
     \begin{subfigure}[b]{8.4cm}
         \centering
         \includegraphics[width=8.4cm]{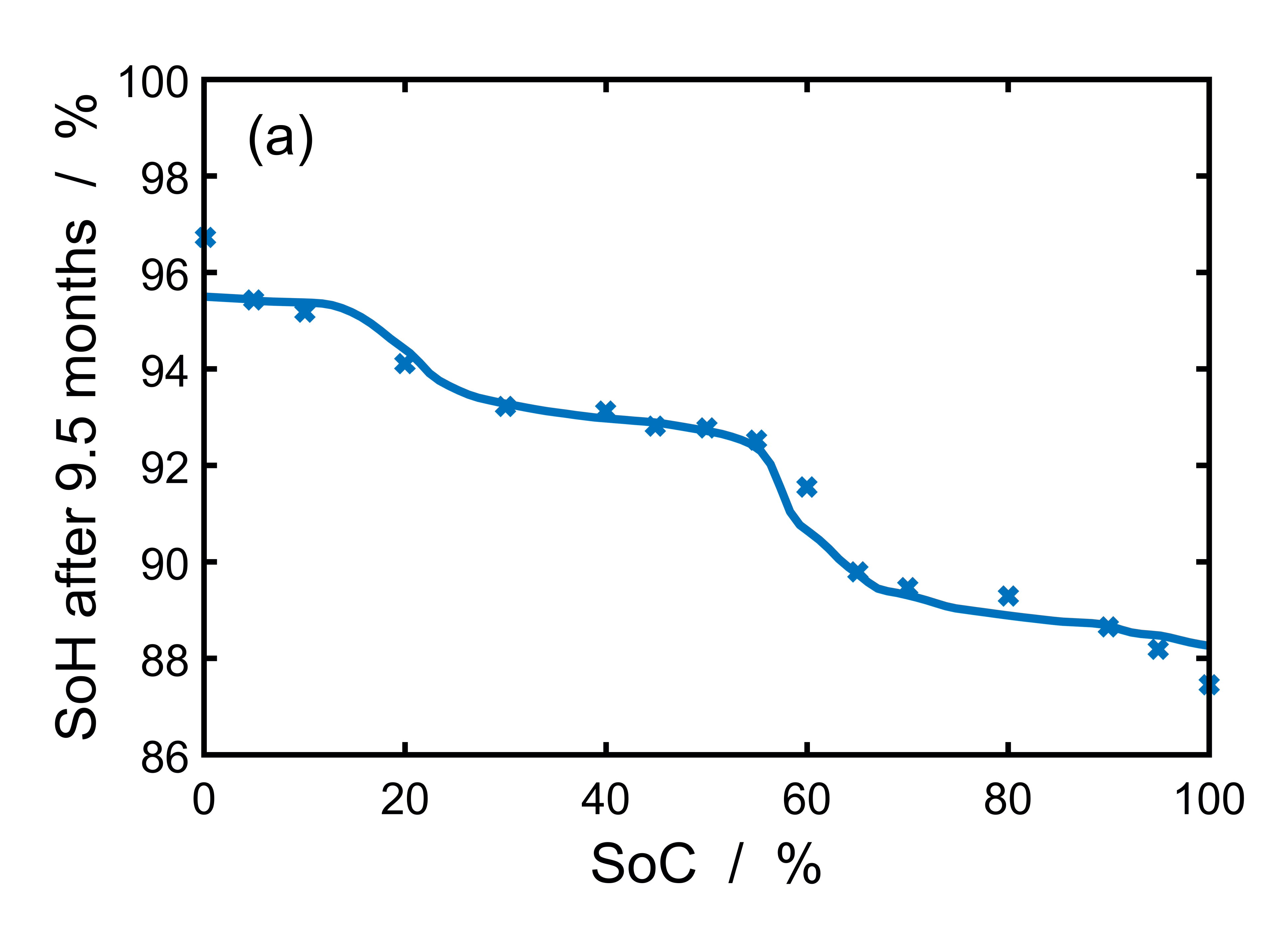}
     \end{subfigure}
     \hfill
     \begin{subfigure}[b]{8.4cm}
         \centering
         \includegraphics[width=8.4cm]{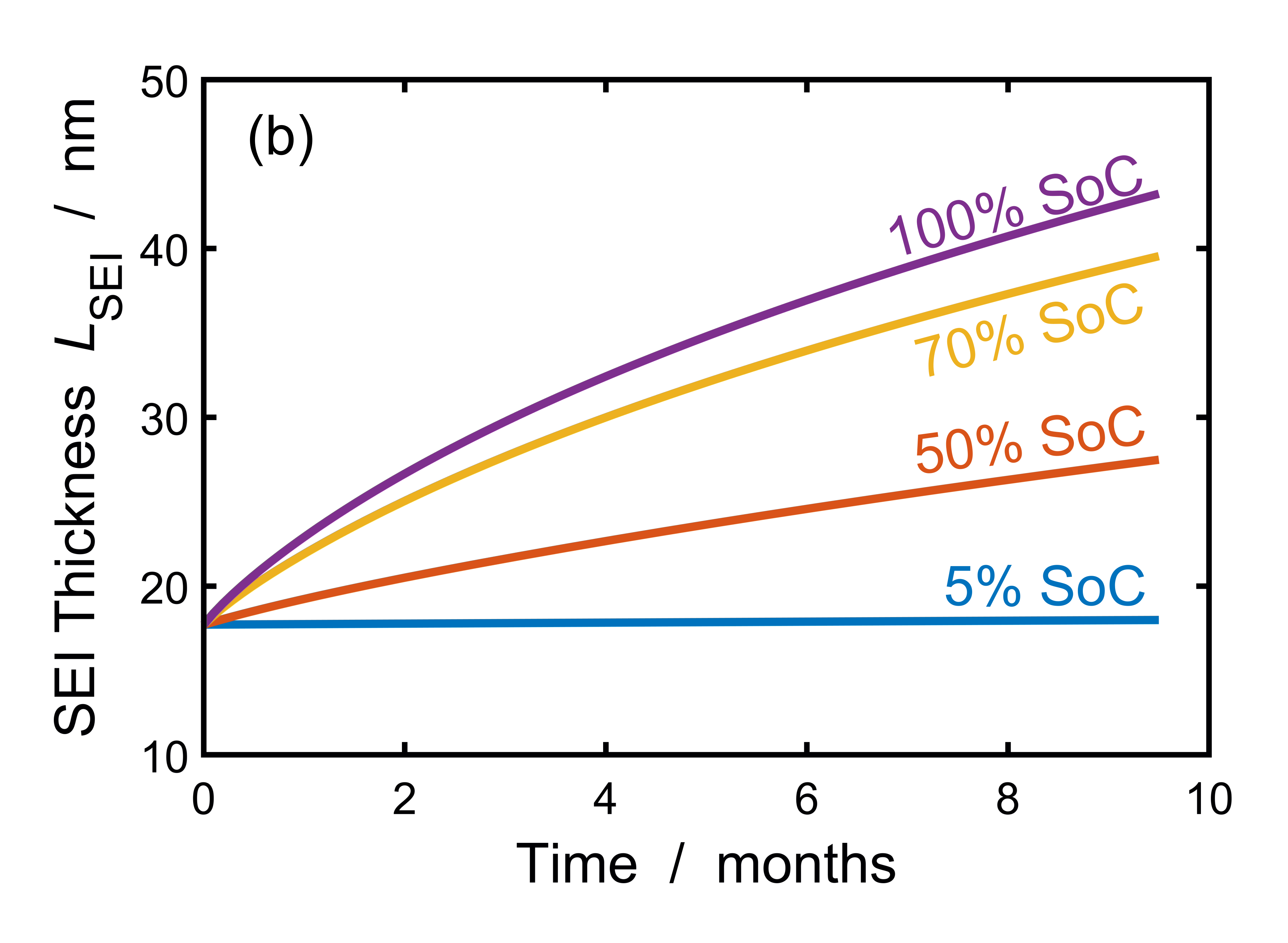}
     \end{subfigure}
        \caption{Parametrization of aging model with SEI growth on graphite electrodes. a) Comparison of the experimental aging data \cite{Keil2016} (crosses) with the simulated results for a graphite anode (line). b) SEI growth over time for different states of charge.}
        \label{fig:Chemical_Validation}
\end{figure}

In Figure \ref{fig:Chemical_Validation} we compare the results of our simulation with the experimental storage data of Keil \textit{et al.} \cite{Keil2016}.
We conclude that our model results in square-root-of-time growth during storage and describes the experimentally observed SoC-dependence well. Based on these growth parameters we now proceed to validate our mechanical SEI model.

\subsubsection*{SEI Mechanics}

We rely on recent membrane bulge measurements to validate our mechanical SEI model \cite{Yoon2020a}. Yoon \textit{et al.} \cite{Yoon2020a} grow an SEI from a thin lithium film located on a polymeric support. By applying pressure, the resulting SEI/polymer film bulges. The pressure/bulge characteristics are then translated to stress-strain curves for the SEI in the circumferential direction. Moreover, atomic force microscopy visualizes cracks inside the SEI depending on its expansion. To mimic these experiments, we expand the SEI continuously with a constant velocity $\dot{r}_1$ at the innermost SEI element and calculate the mean circumferential SEI stress $\bar{\sigma}_{\SEI,\phi}=1/L_\SEI \int {\sigma}_{\SEI,\phi} \text{d}r$ and expansion $\bar{\epsilon}_{\SEI,\phi}=1/L_\SEI \int \tens{F}_{\SEI,\phi}-1 \text{d}r$.

\begin{figure}[tb] 
 \centering
 \includegraphics[width=8.4 cm]{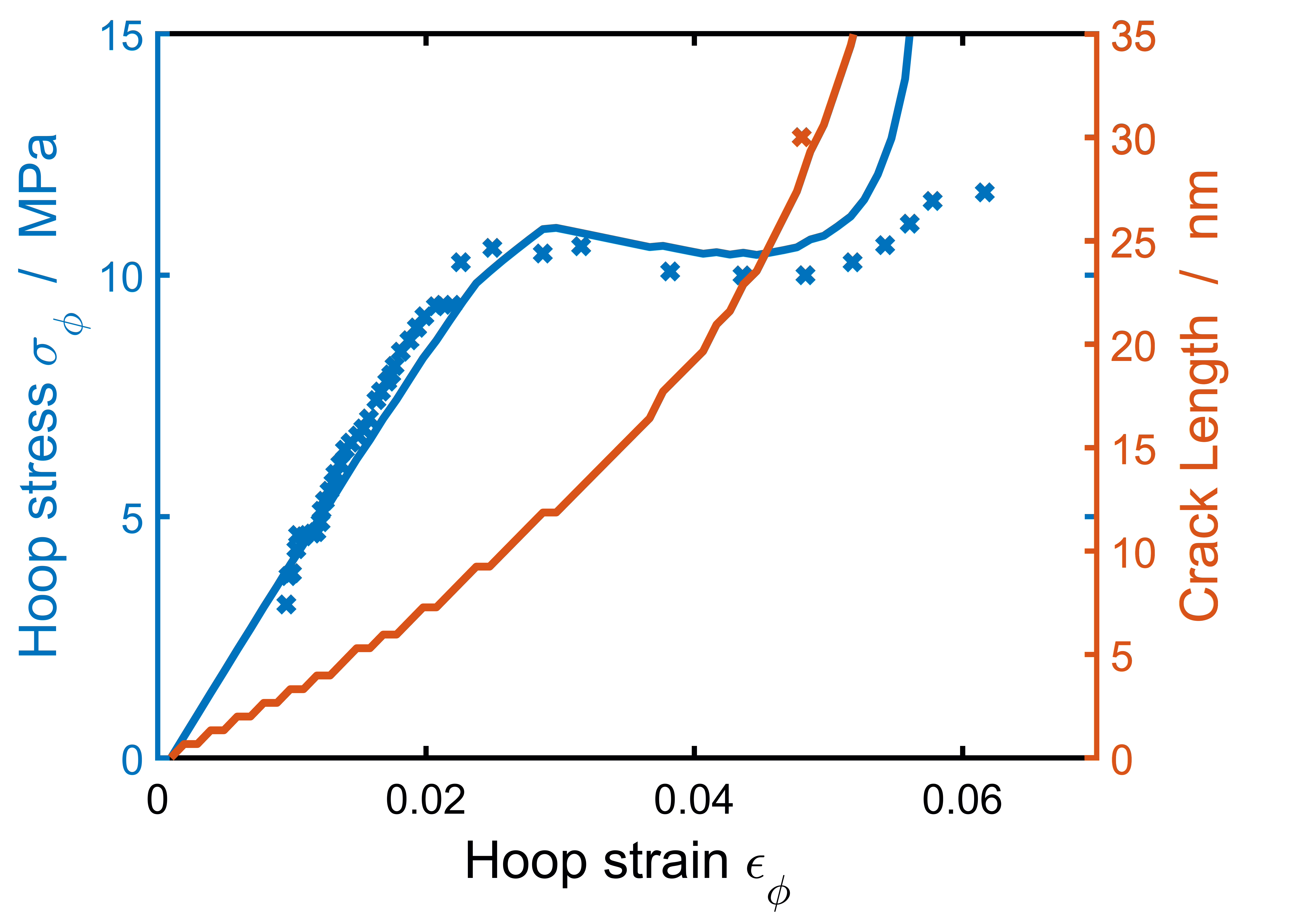}
 \caption{Comparison of mechanical experiments on SEI from lithium thin films (crosses) \cite{Yoon2020a} with the simulated results (line) to parametrize the mechanical SEI model. Stress-strain curve in blue, SEI crack formation in orange. The steps in both curves result from the numerical discretization of the SEI.}
 \label{fig:Mechanical_Validation}
\end{figure}

Figure \ref{fig:Mechanical_Validation} shows the experimental results of Yoon \textit{et al.} \cite{Yoon2020a} compared to our simulation results.
We see that our model agrees well with the experimental stress-strain curve. Furthermore, our SEI fracture model matches the experimentally observed crack evolution.

\section{Results and Validation}

In this section, we analyze the model outlined above. We start by studying the particle-SEI geometry during one cycle in section \ref{ss:Geometry}. Next, we take a closer look at the mechanical response of silicon particle and SEI in section \ref{ss:Mechanics}. Subsequently, we analyze the SEI growth during one cycle in section \ref{ss:ShortTermGrowth} and finally look at the long-term SEI growth \ref{ss:LongTermGrowth}. Unless otherwise specified, the particles were cycled at 1C between $U_{\text{max}}=\SI{0.5}{\volt}$ and $U_{\text{min}}=\SI{0.05}{\volt}$.

\subsection{Geometry}
\label{ss:Geometry}
\begin{figure}[tb] 
 \centering
 \includegraphics[width=8.4 cm]{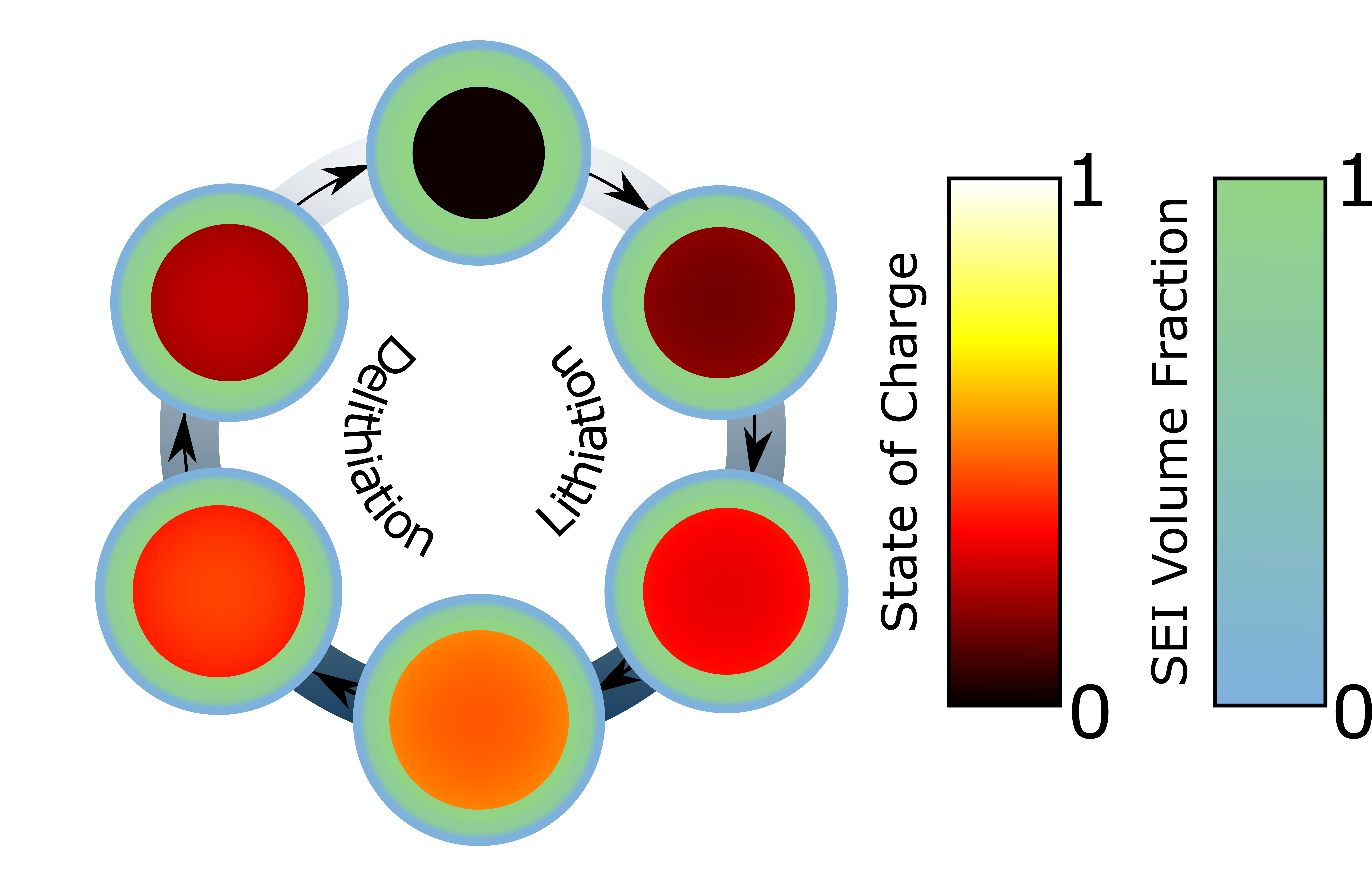}
 \caption{Geometrical representation of our spherical symmetric simulation setup, which consists of a spherical silicon particle surrounded by a porous SEI. Varying state of charge during cycling induces changes of particle size and SEI morphology.}
 \label{fig:Geometry_during_Cycling}
\end{figure}

Lithiation and delithiation strongly affect particle and SEI geometry. In Figure \ref{fig:Geometry_during_Cycling}, we show six distinct configurations of our spherical symmetric simulation domain during a battery cycle.
We see that the varying state of charge induces volume changes inside the electrode particle according to the chemical expansion $\tens{F}_\ch$. The surrounding SEI responds to this volume change by thinning for high SoC and thickening for low SoC. 

\begin{figure}[tb]
     \centering
     \begin{subfigure}[b]{8.4cm}
         \centering
         \includegraphics[width=8.4cm]{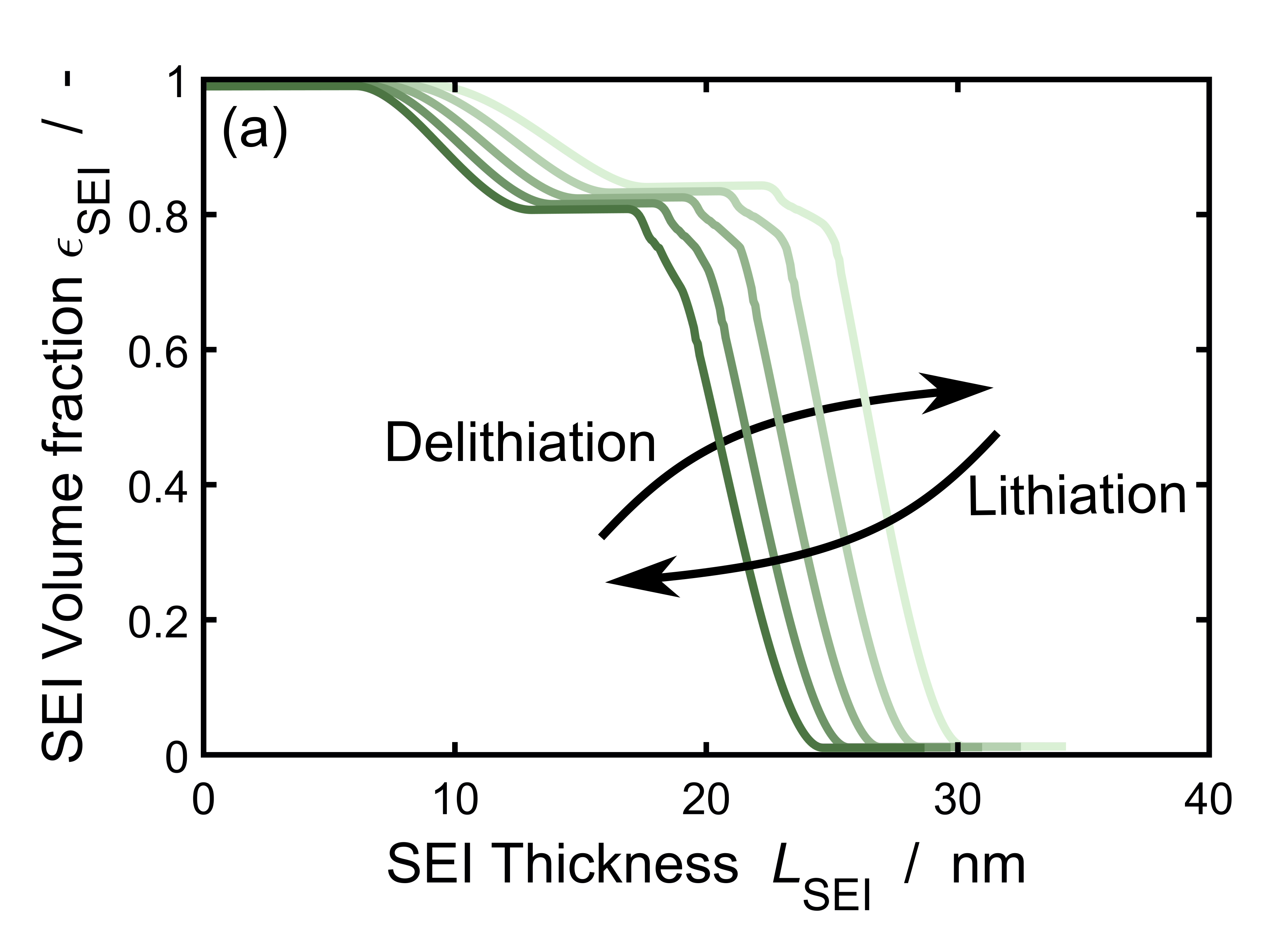}
     \end{subfigure}     
     \hfill
     \begin{subfigure}[b]{8.4cm}
         \centering
         \includegraphics[width=8.4cm]{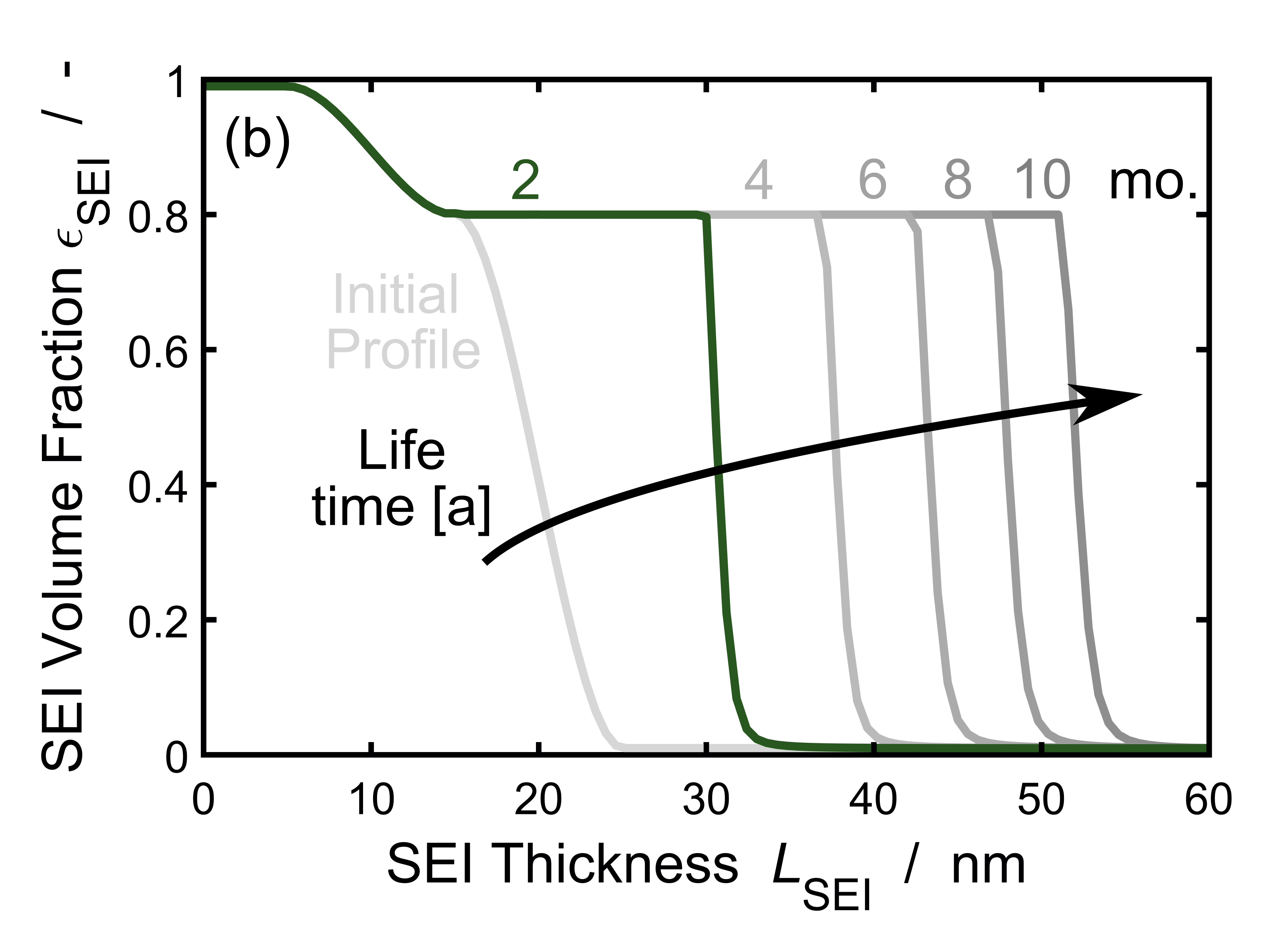}
     \end{subfigure}
 \caption{SEI porosity evolution over time. a) Reversible mechanical breathing of the SEI during one cycle. b) Irreversible chemical SEI growth over several months during storage.}
 \label{fig:Breathing_during_Cycle}
\end{figure}

We further resolve the geometrical response of the SEI in Figure \ref{fig:Breathing_during_Cycle}.
In Figure \ref{fig:Breathing_during_Cycle} a), we see the SEI breathing in each cycle. During delithiation, the SEI expands from $\SI{20}{\nano\meter}$ to $\SI{30}{\nano\meter}$. Moreover, the SEI densifies during delithiation leading to a higher SEI volume fraction.
The SEI expansion and densification result from the volume conservation of the SEI matrix. As the electrode particle shrinks, so does the inner radius of the SEI shell. The SEI compensates this loss in volume by increasing its volume fraction and thickness. 
This reversible short-term expansion of the SEI overshadows the long-term SEI growth taking place in a time scale of months.
In Figure \ref{fig:Breathing_during_Cycle}, we thus isolate this long-term change in geometry by plotting the SEI thickness during storage over several months.  We observe, that the initial profile grows into a dual layer SEI structure with a sharp front to the electrolyte. The dual layer structure agrees with the predictions of Single \textit{et al.} \cite{Single2016,Single2017} and is enforced here via the limiting porosity $\epsilon_{\ely,\text{min}}$. Our choice of a fast reaction rate, verified by the experiments in Figure \ref{fig:Chemical_Validation} yields the sharp reaction front.
 
\subsection{Mechanics} 
\label{ss:Mechanics}

\begin{figure*}[htb] 
 \centering
 \includegraphics[width=17.8 cm]{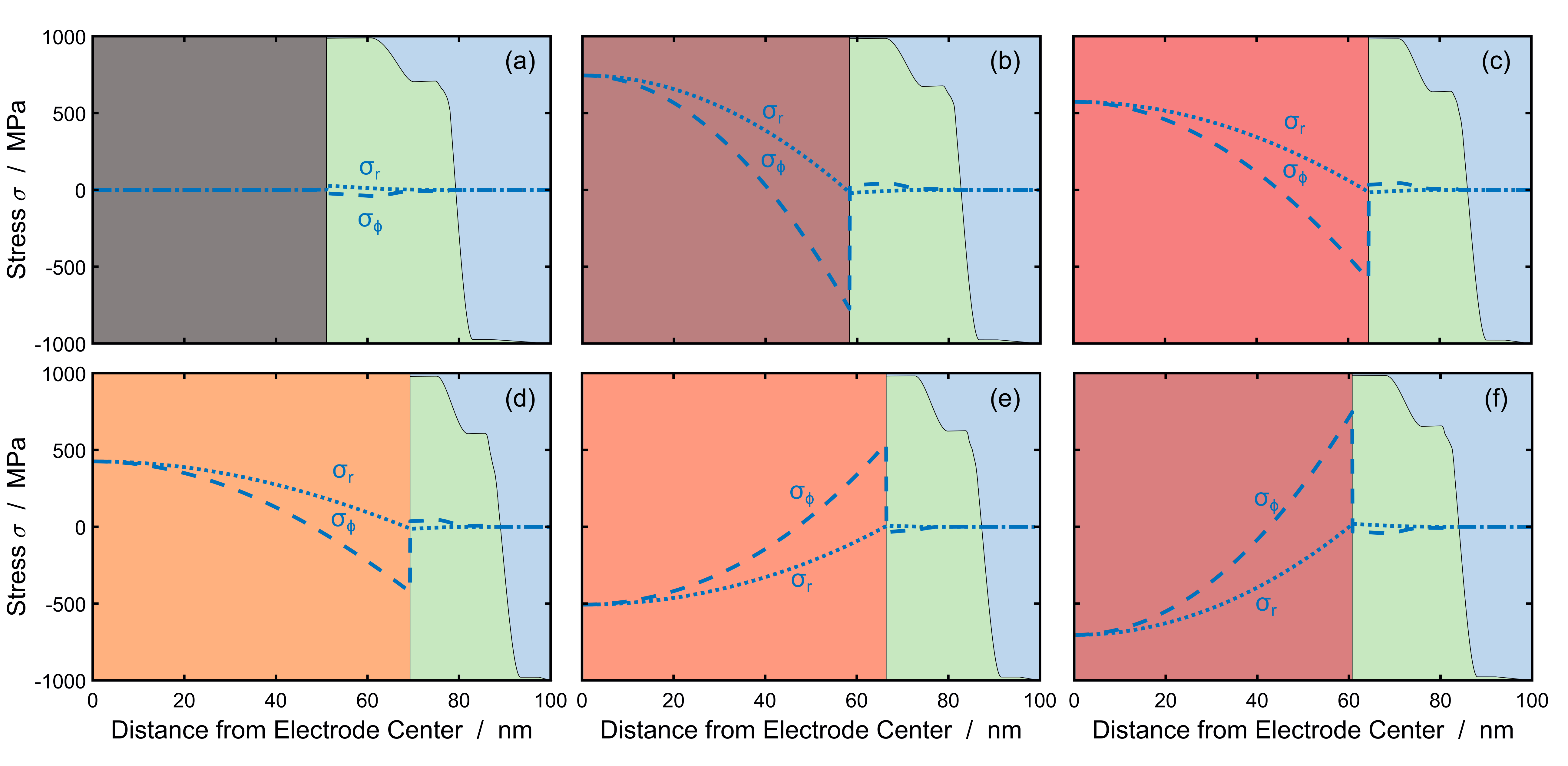}
 \caption{Radial stress $\sigma_r$ (dotted) and hoop stress $\sigma_\phi$ (dashed) inside particle and SEI during one cycle. The six snapshots correspond to the six geometries shown in Figure \ref{fig:Geometry_during_Cycling}. In terms of color, the electrode domain has the respective SoC color (see Figure \ref{fig:Geometry_during_Cycling}) and the SEI/electrolyte domain is subdivided into SEI (green) and electrolyte (blue) according to the porosity profile.}
 \label{fig:Stress_during_Cycle}
\end{figure*}

Next, we investigate the mechanical response of particle and SEI to the previously discussed geometrical changes. Figure \ref{fig:Stress_during_Cycle} shows the stress state in the six different configurations of Figure \ref{fig:Geometry_during_Cycling}.
Initially, in Figure \ref{fig:Stress_during_Cycle} a), the electrode particle is stress free. The lithiation half-cycle, Figure \ref{fig:Stress_during_Cycle} b)-d), induces tensile stress in the inner part of the particle and compressive stress in the outer part. During delithiation, Figure \ref{fig:Stress_during_Cycle} e)-f), this behavior is inverted with tensile stress in the inner part and compressive stress in the outer part.
The stress inside the SEI in contrast is fully compressive during lithiation and fully tensile during delithiation. We observe two interesting phenomena in the stress response of the SEI.
First, the SEI is initially not stress free. Second, the stress magnitude inside the SEI remains largely constant over the cycle and differs largely between inner and outer SEI.

The stress profile inside the particle stems from concentration gradients. Initially, the concentration is homogeneous so that the stress vanishes. Upon lithiation, the concentration in the particle center is lower than at its surface leading to volume mismatches. To compensate this mismatch, the particle center stretches elastically while the outer particle compresses elastically. This causes the observed tensile stress in the particle center and the compressive stress in the outer particle. During charging, the concentration gradient is inverted leading to the opposite behavior \cite{Castelli2021}.

The SEI stress response in contrast is solely caused by the particle breathing, because the lithium ion concentration inside the SEI is constant \cite{Single2019}.
The initial SEI stress in Figure \ref{fig:Geometry_during_Cycling} a) results from the SEI deformation from its reference configuration at $U_\rf=\SI{0.15}{\volt}$ \cite{Tokranov2014,Tokranov2016,Kumar2016,Veith2017} to $U_\text{min}=\SI{0.05}{\volt}$. Subsequently, the stress magnitude inside the SEI remains largely constant due to plastic deformation. The observed stress is thus the yield causing stress with $f(\sigma)=0$, see Equation \ref{eq:yield_function_SEI} and Figure \ref{fig:Yield_Surface}. Along the SEI, we observe a stress profile due to the prescribed dual layer structure, see Equation \ref{eq:Transition}. The lower limiting porosity $\epsilon_{\ely,\text{min}}$ and the higher Young's modulus $E_\SEI$ and yield strength $\sigma_\text{Y}$ of the dense, inorganic inner layer lead to a higher stress magnitude compared to the porous, organic outer layer.

\begin{figure}[tb] 
 \centering
 \includegraphics[width=8.4 cm]{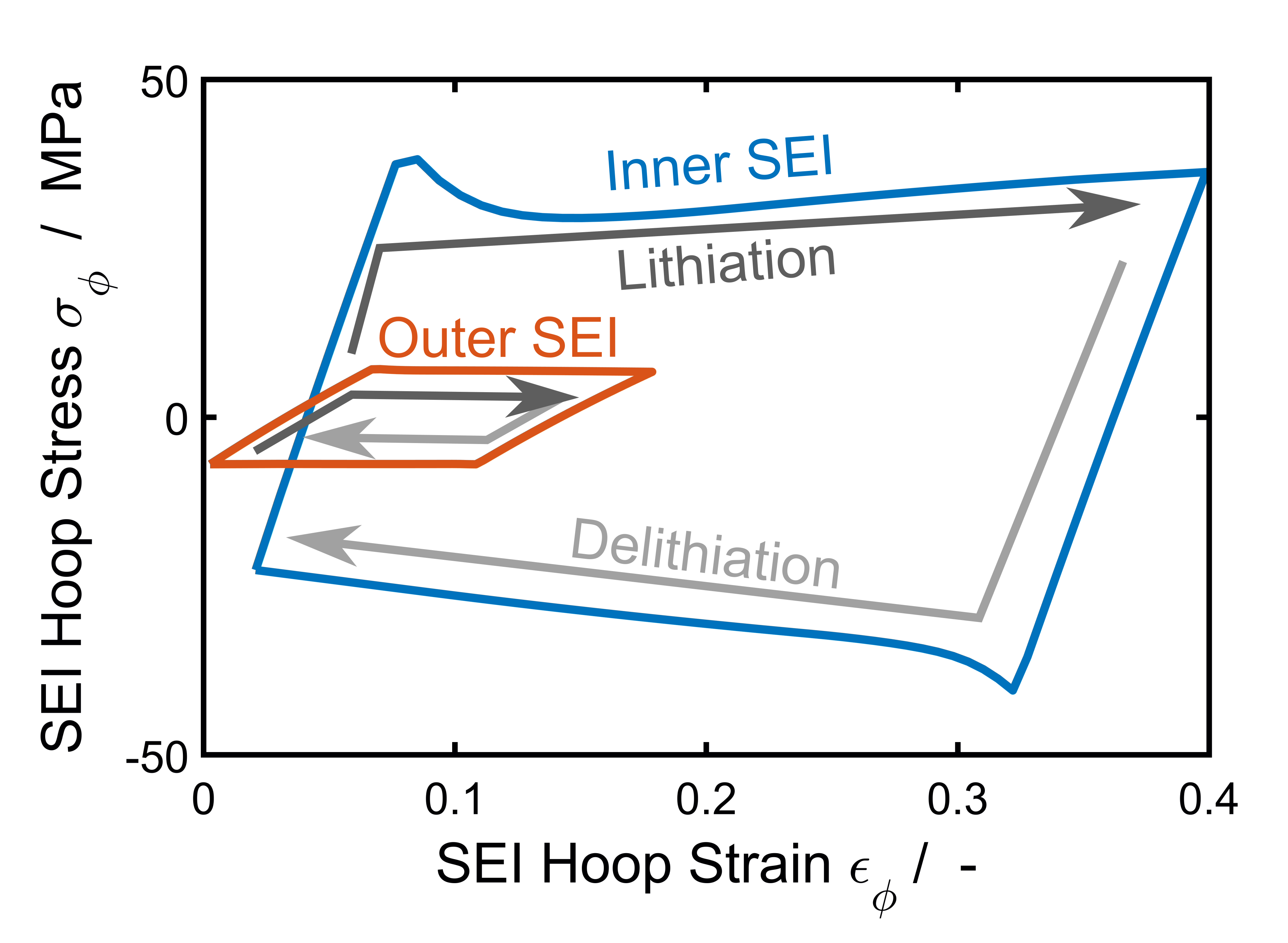}
 \caption{Stress strain curves for the inner (blue) and outer (orange) SEI during cycling. 
 The SEI first deforms linear elastic and then flows perfectly plastic upon reaching the yield stress. Lithiation and delithiation induce opposing mechanical loads, which causes the observed hysteresis during one cycle.}
 \label{fig:Hysteresis}
\end{figure}

We further analyze the mechanical response of the SEI by plotting the stress-strain curve inside the inner and the outer SEI in Figure \ref{fig:Hysteresis}. 
We observe a hysteresis in the stress response of the SEI with tensile stress during lithiation and compressive stress during delithiation. During lithiation, the SEI expands linear elastically in hoop direction until it reaches an expansion of $5\%$. Then the plastic deformation sets in and expands the inner SEI as much as $40\%$ and the outer SEI around $20\%$. The stress magnitude in the outer SEI is approximately constant at $\SI{10}{\mega\pascal}$, because we assume perfect plasticity \cite{Yoon2020a}. In contrast, the stress magnitude in the inner SEI is not as constant, but varies around a value of $\SI{40}{\mega\pascal}$. This difference between inner and outer SEI results from the radial stress component, see Figure \ref{fig:Stress_during_Cycle}. While the radial stress in the inner SEI varies during the cycle due to the mechanical particle-SEI boundary condition the radial stress in the outer SEI is negligible. During delithiation, the tensile stress releases elastically until a compression of $5\%$ is reached. Then the plastic flow compresses the SEI in hoop direction with a similar stress and strain magnitude as in the lithiation half-cycle. 

So far, we analyzed elastic and plastic SEI deformation and observed no SEI fracture. This is because our model SEI was formed at a largely expanded particle with $U_\rf=\SI{0.15}{\volt}$. Thereby, the SEI exhibits large compressive, but only small tensile strains in hoop direction, which effectively prevents SEI fracture. Nevertheless, large compressions might lead to other mechanical failure modes like buckling or delamination, which we do not consider in our reductionist model \cite{Kumar2017,Guo2020}.
To further analyze SEI fracture, we thus subject our SEI to larger tensile strains. We therefore set the stress-free SEI configuration to a smaller particle size with $U_\rf=\SI{0.3}{\volt}$ and cycle with C/100 to increase the SoC swing. 

\begin{figure}[tb] 
 \centering
 \includegraphics[width=8.4 cm]{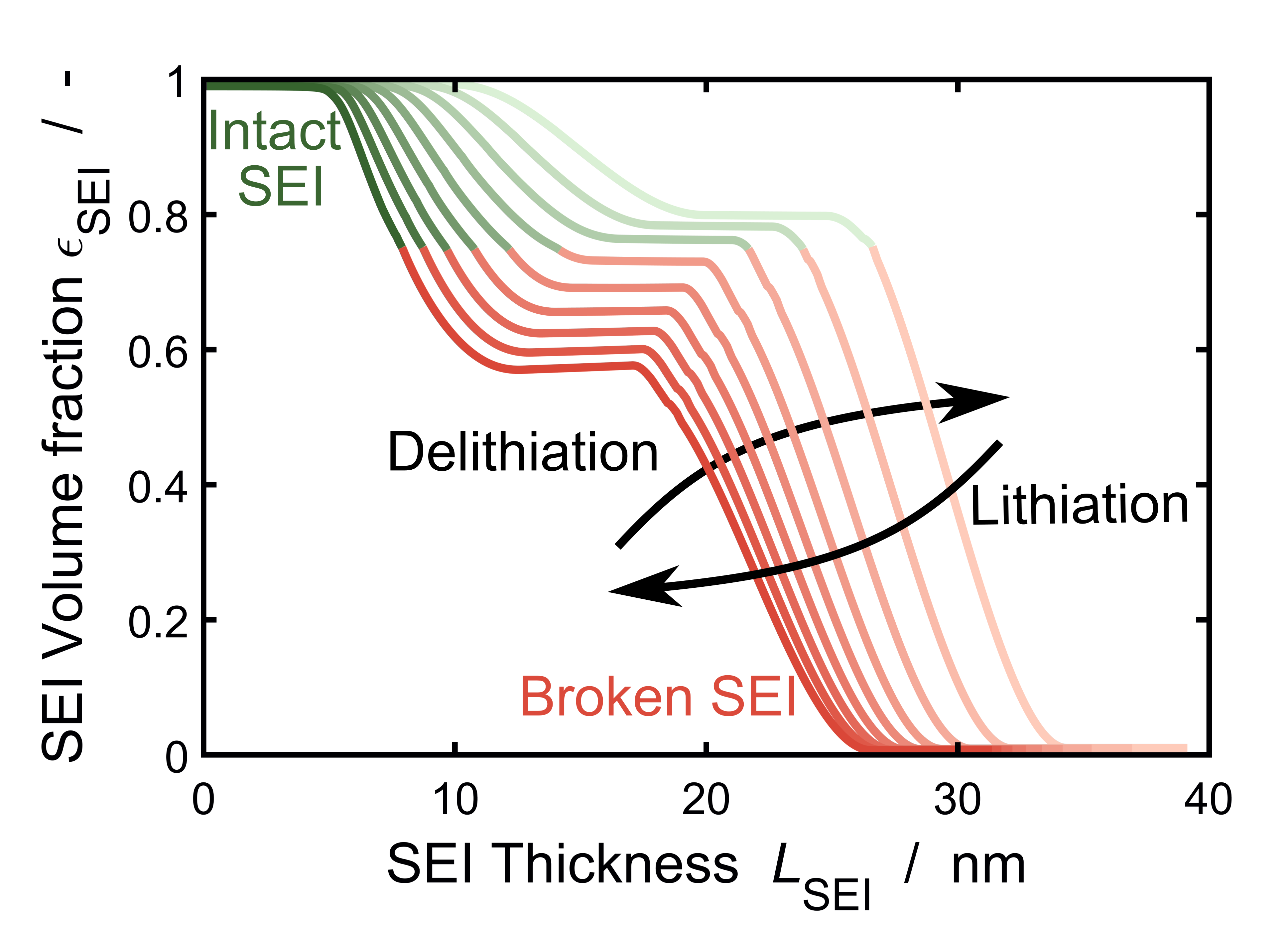}
 \caption{SEI porosity profile evolution under large mechanical stress. The SEI consists of intact (green) and broken (red) domains. Darker colors indicate higher particle SoC. Accordingly, the mechanical SEI degradation increases with the particle state of charge.}
 \label{fig:SEIFracture}
\end{figure}

With $U_\rf=\SI{0.3}{\volt}$, we can now study SEI fraction within a single cycle. Figure \ref{fig:SEIFracture} shows the proceeding SEI fracture for this simulation setup during lithiation. 
The crack starts from the SEI surface and expands through the outer SEI stopping as it approaches the inner SEI. We observe that the SEI deteriorates much stronger once it is broken leading to larger pore expansion compared to our standard cycle shown in Figure \ref{fig:Breathing_during_Cycle}. However, similar to the fully intact case in Figure \ref{fig:Breathing_during_Cycle}, we see that this deformation is reversible and the crack closes again as the particle delithiates and the SEI compresses. This accords well to the experimental findings of Kumar \textit{et al.} \cite{Kumar2016}, who observed SEI cracks only in the outer SEI, which close again upon delithiation. However, our homogenized 1D model cannot capture the precise shape of the cracks and whether the same cracks would open again in the next cycle \cite{Kumar2016}.
Overall, this mechanism accelerates SEI growth by lowering the SEI thickness and increasing the pore volume. These results show that low potentials in the initial SEI formation cycle increase the battery lifetime by enhancing the mechanical stability of the SEI. In the next two sections, we analyze the SEI growth during short-term and long-term cycling.

\subsection{Short-Term SEI Growth}
\label{ss:ShortTermGrowth}

\begin{figure}[tb] 
 \centering
 \includegraphics[width=8.4 cm]{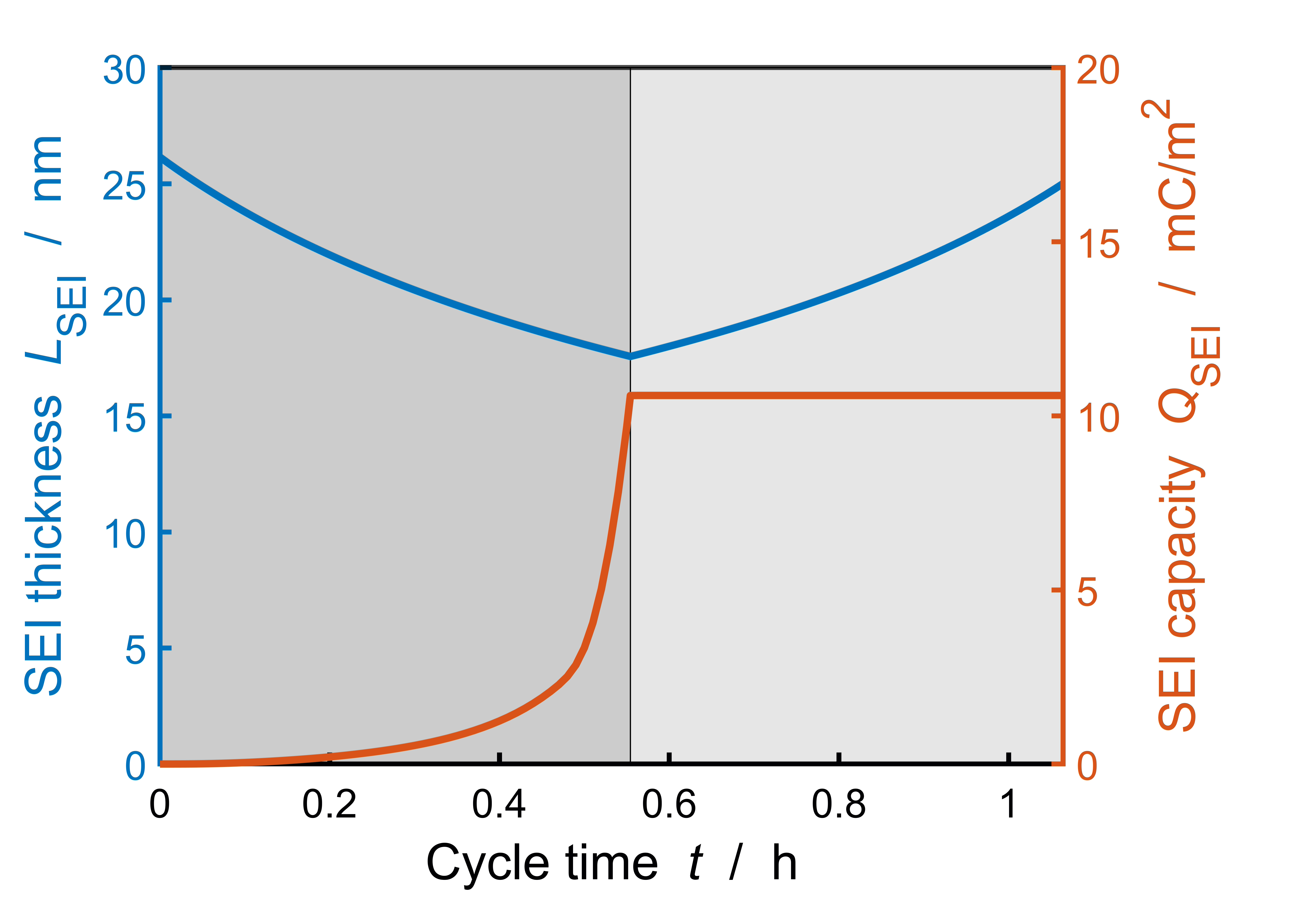}
 \caption{SEI thickness (blue) and capacity (orange) during charging (dark grey) and discharging (light grey) with 1C. The thickness shows the reversible mechanical SEI breathing during each cycle. The SEI capacity elucidates the asymmetric SEI growth, which is accelerated during lithiation and decelerated during discharging.}
 \label{fig:Asymmetry}
\end{figure}

Significant SEI growth typically occurs in a time span of months to years. Nevertheless, our model allows us to visualize the small SEI growth during one cycle.
We start by analyzing the SEI thickness $L_\SEI=R(\epsilon_\SEI>0.05)$ and the SEI capacity consumption $Q_\SEI=2F/\bar{V}_\SEI\int \varepsilon_\SEI 4\pi R^2 \text{d}R$ during our standard cycle in Figure \ref{fig:Asymmetry}.
We observe a reversible thinning and thickening of the SEI during the cycle, corresponding to our findings from Figure \ref{fig:Breathing_during_Cycle}. The irreversible SEI growth only plays a minor role, so that the initial thickness approximately corresponds to the final thickness. We thus resolve the irreversibly consumed SEI capacity $Q_\SEI$ during the cycle on the second y-axis and observe asymmetric capacity consumption during one cycle: Irreversible SEI growth is accelerated by lithiation and decelerated by delithiation. Moreover, the SEI growth is fastest at the end of lithiation, \textit{i.e.}, for high SoC. 

The asymmetric SEI growth results from the exponential dependence of the lithium interstitial concentration on the anode OCV $U_0$ and intercalation overpotential $\eta_\text{int}$ \cite{VonKolzenberg2020}. 
The influence of anode OCV on SEI growth was first theoretically described by Single \textit{et al.} \cite{Single2018} in line with the storage experiments of Keil \textit{et al.} \cite{Keil2016}, see Figure \ref{fig:Chemical_Validation}. The influence of intercalation overpotential stems from our recent electrochemical SEI model \cite{VonKolzenberg2020}. This model agrees well to the experiments of Attia \textit{et al.} \cite{Attia2019}, which revealed dependence of SEI growth on the current magnitude and direction.

\subsection{Long-Term SEI Growth}
\label{ss:LongTermGrowth}

\begin{figure}[tb!]
     \begin{subfigure}[b]{8.4cm}
         \centering
         \includegraphics[width=8.4cm]{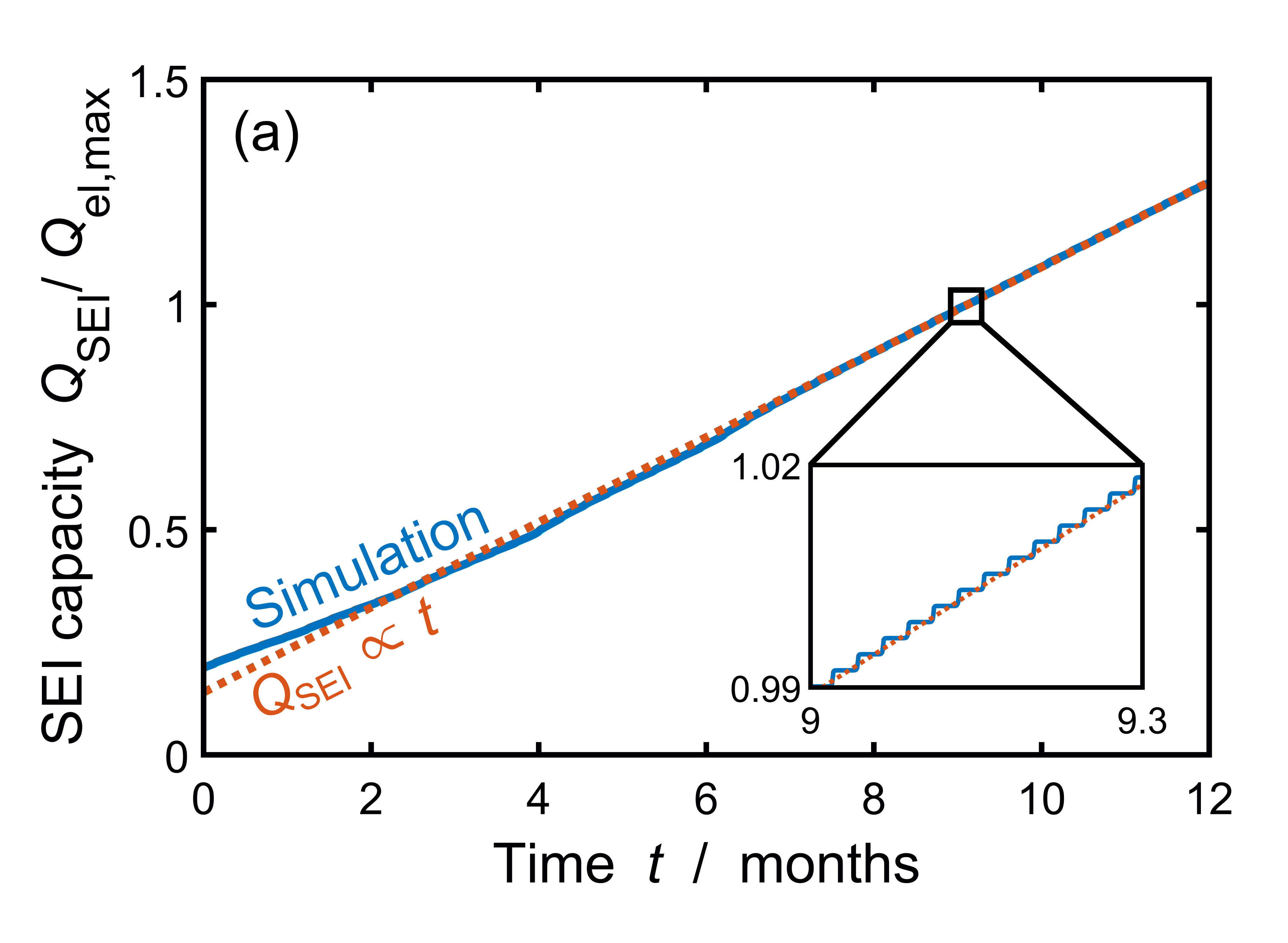}
     \end{subfigure}\\    
     \begin{subfigure}[b]{8.4cm}
         \centering
         \includegraphics[width=8.4cm]{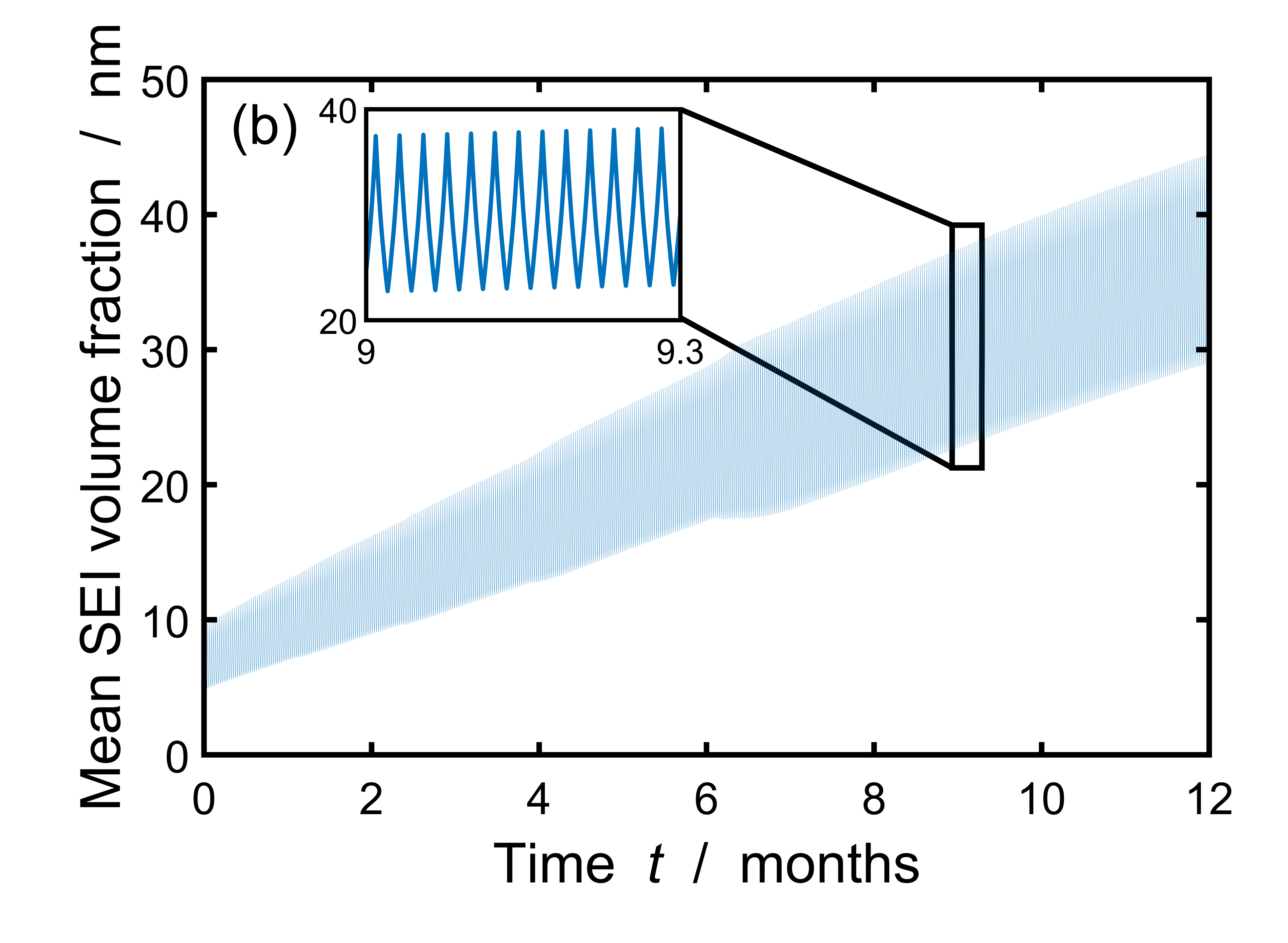}
     \end{subfigure}
 \caption{SEI mechanics and growth on a silicon electrode during one year of continuous cycling with C/10. (a) Irreversibly bound SEI capacity relative to electrode capacity $Q_\SEI/Q_\text{el,max}$ (blue) compared to a linear growth law (dashed orange line). The zoom inlet shows the asymmetric growth in each cycle, see Figure \ref{fig:Asymmetry}. (b) Evolution of mean SEI volume fraction $\bar{\epsilon}_\SEI=\int \epsilon_\SEI \text{d}r$.}
 \label{fig:Oscillation}
\end{figure}

Because the SEI hardly grows during a single cycle, we analyze the long-term SEI growth after several battery cycles in this subsection. In Figure \ref{fig:Oscillation} a), we illustrate the capacity $Q_\SEI$ bound in the SEI of a silicon particle cycled for one year at C/10 with the standard potential of initial SEI formation $U_\rf=\SI{0.15}{\volt}$. Additionally, we plot the mean SEI volume fraction $\bar{\epsilon}_\SEI=\int \epsilon_\SEI \text{d}r$ in Figure \ref{fig:Oscillation} b). This quantity enriches our analysis of morphological SEI changes, because it also captures the influence of reversible SEI densification/porosification during each cycle, which we observed in Figure \ref{fig:Breathing_during_Cycle} and \ref{fig:SEIFracture}. Moreover, our SEI growth model, Equation \ref{eq:Porosity_Change_Differential}, rather depends on the porosity profile $\epsilon_\ely(R)$ than the macroscopic SEI thickness $L_\SEI$.

In Figure \ref{fig:Oscillation} a), we observe the same trend as in Figure \ref{fig:Asymmetry}
during each cycle (see zoom inlet); the SEI thickness oscillates and the SEI capacity grows asymmetrically.
Similarly, the mean SEI volume fraction $\bar{\epsilon}_\SEI$ oscillates, shown in the zoom inlet of Figure \ref{fig:Oscillation} b).
Over the long-term, the amplitude of oscillations of mean SEI volume $\bar{\epsilon}_\SEI$ increases from $\SI{5}{\nano\meter}$ to $\SI{15}{\nano\meter}$ as SEI fracture progresses, see Figure \ref{fig:SEIFracture}. 
The fracture in turn decreases the SEI passivation so that we observe a linear capacity fade in Figure \ref{fig:Oscillation} a). 
In contrast, during storage, we observe a self-limiting SEI growth $Q_\SEI \propto \sqrt{t}$, see Figure \ref{fig:Chemical_Validation} b).
This accelerated growth results from the interplay of battery cycling and SEI growth, i.e., from mechanical SEI deterioration. 

\begin{figure}[tb] 
 \centering
 \includegraphics[width=8.4cm]{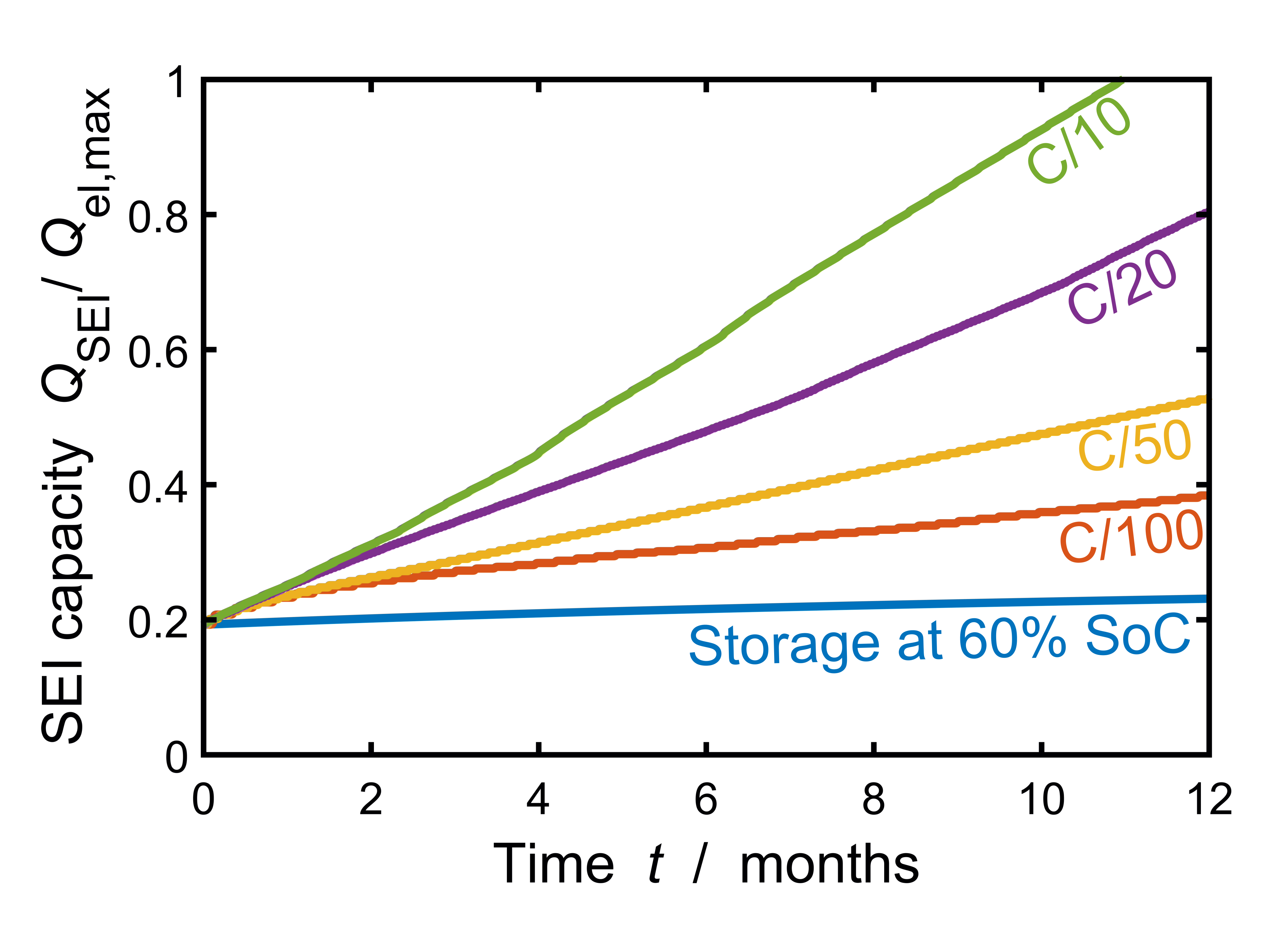}
 \caption{Irreversible SEI capacity consumption $Q_\SEI$ relative to the maximum particle capacity $Q_{\text{el,max}}$ for silicon particles cycled for one year with different charging currents.}
 \label{fig:SEIGrowthCycling}
\end{figure}

In Figure \ref{fig:SEIGrowthCycling}, we thus illustrate how the SEI capacity increases over time for different charging currents as compared to battery storage. 
We observe that higher charging currents lead to faster SEI growth. But the relationship between charging current and SEI growth seems to be more complex, as we observe only a small difference between C/100 and C/50 opposed to the large difference between C/50 and C/20.

\begin{figure}[tb] 
 \centering
 \includegraphics[width=8.4cm]{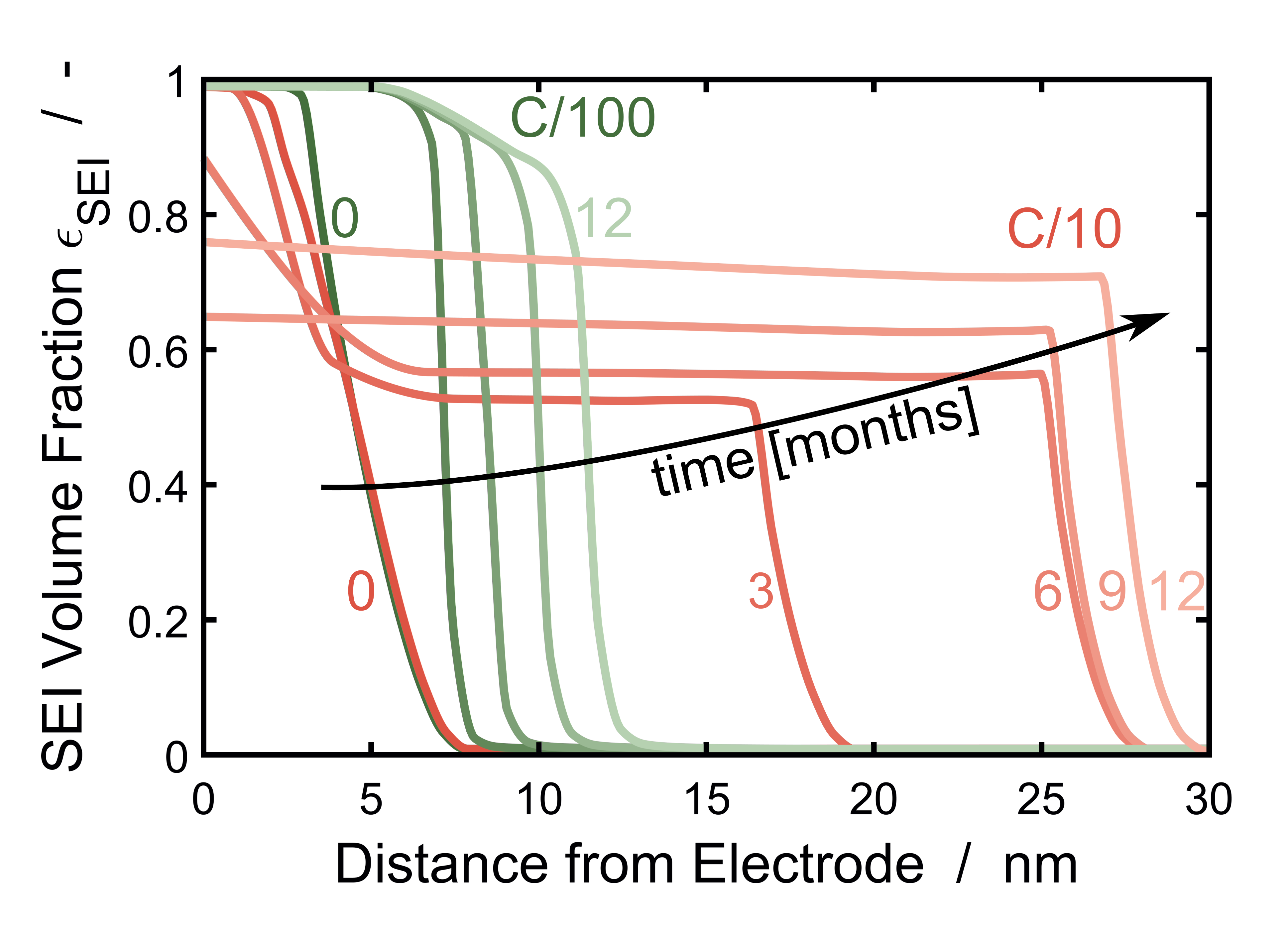}
 \caption{Evolution of the SEI volume fraction profile on a silicon particle cycled with C/100 (green) or C/10 (red) for one year.}
 \label{fig:PorosityEvolution}
\end{figure}

In Figure \ref{fig:PorosityEvolution}, we therefore plot the dynamic SEI profile for a particle charged with C/100 (green) and C/10 (red). 
We observe a fundamentally different SEI growth between these two charging currents. The particle charged with C/100 closely follows the prescribed limiting profile similar to the storage case, Figure \ref{fig:Breathing_during_Cycle} b). In contrast, charging with C/10 leads to a spread out profile in which the shielding inner layer vanishes over time.

The reason for this difference lies in the different time-scales imposed by the different charging rates. In each cycle, the inner SEI undergoes large deformations as shown in Figure \ref{fig:Breathing_during_Cycle} a). 
These large deformations are not completely reversible and the inner SEI layer flows plastically to a thicker and more porous geometry. If the battery cycles with C/100, the cycle time is sufficient for the inner SEI to reform in the newly created pores and thereby reattain its self-passivating character. However, in batteries charged with C/10, the cycle time is too short for the SEI to reform. As a result, the inner SEI fully deteriorates and leaves the anode unshielded from the electrolyte. This causes unlimited SEI growth leading to the observed linear growth in the long-term.

To sum up our long-term results for SEI growth shown in Figure \ref{fig:Oscillation}, \ref{fig:SEIGrowthCycling}, and \ref{fig:PorosityEvolution}, we observe a fundamental transition in time-dependence. Starting from the well-known $\sqrt{t}$-SEI-growth during storage, cycling with increasing current accelerates growth and at C/10 the SEI grows linear with time.
This growth acceleration stems from the continuous pore expansion inside the inner SEI caused by large deformations during cycling. The faster the cycling rate, the less time the SEI has to repair these pores. As a result, the porosity profile in Figure \ref{fig:PorosityEvolution} at C/100 corresponds to the prescribed SEI profile, which we also observe during battery storage, see Figure \ref{fig:Chemical_Validation} and \ref{fig:SEIFracture}. In contrast, faster cycling with C/10 deteriorates the inner SEI over time, so that the particle is no longer passivated and the SEI grows rapidly. 
This finding rationalizes empirically motivated SEI growth models, which obtain a linear SEI growth from prescribing a constant SEI fracture and regrowth term for every cycle \cite{Pinson2012,Li2015,Perassi2019}.

Despite the qualitatively good predictions, our model is only a first step towards better understanding SEI growth on silicon electrode particles. 
First, our parametrization with SEI experiments on graphite \cite{Keil2016} and lithium thin film \cite{Yoon2020a} neglects structural differences of the SEI on the different substrates and thereby leads to uncertainties. Novel capacity fade experiments on silicon particles will be helpful to obtain better estimates for the model parameters. 
Second, we rely on a 1D-spherical symmetrical single particle model instead of resolving the whole electrode.
As such, our model is agnostic of the circumferentially heterogeneous SEI structure and anisotropy within the electrode particles. 
Third, we only consider a single particle instead of analyzing the electrode microstructure as a whole. Thereby, our model neglects structural heterogeneity resulting from particle size distribution, as well as graphite, carbon and binder phases. This heterogeneity leads to complex electrochemo-mechanical interactions, especially in graphite/silicon blend electrodes as recently shown by Liu \textit{et al.} \cite{Liu2021}.

However, these limitations offer promising possibilities to extend and apply our model. 
With efficient numerical schemes, \textit{e.g.}, as presented  in ref. \citenum{Castelli2021}, our model can be analyzed in 3D to reveal heterogeneity of SEI growth on individual particles. Moreover, incorporating a simplified 0D-version of our model in 3D-resolved electrode models \cite{Latz2011,Latz2015,Liu2021} is a promising approach to reveal interparticular SEI heterogeneity as well as the complex electrochemo-mechanical interactions of the different constituents. In this context, our SEI model can direcly be adapted to describe the mechanics of binder shell covered electrode particles.

\section{Conclusion}

We have developed a thermodynamically consistent chemo-mechanical model of an electrode particle coated with a solid-electrolyte interphase (SEI). The electrode model is derived from a free energy functional and accounts for chemical deformation and elastic stress \cite{Castelli2021}. The SEI model accounts for elastic and plastic deformation, fracture, and lithium atom mediated SEI growth \cite{Single2018,VonKolzenberg2020} based on the SEI volume fraction as order parameter.

Our model agrees qualitatively well with the experimentally observed SEI cracking during lithiation and healing during delithiation on silicon particles \cite{Kumar2016,Guo2020}. We obtain the parametrization for our electrochemo-mechanical SEI model from experiments on different substrates. The storage experiments of Keil \textit{et al.} \cite{Keil2016} conducted on graphite particles parametrize the electrochemical part of the SEI model. The mechanical measurements of Yoon \textit{et al.} \cite{Yoon2020a} conducted on SEI from lithium thin film parametrizes the mechanical part of the SEI model.

For the first time, our so-validated model showed the complex relationship between SEI mechanics and electrochemical growth on silicon electrodes. Namely, mechanical SEI pore expansion further accelerates SEI growth at high states of charge. Moreover, continuous pore creation during SEI expansion deteriorates the inner SEI in the long-term. For cycling currents $J>\text{C}/20$, the cycle time is too short to repair the inner SEI.
As a result, we observe a transition from self-passivating ($\sqrt{t}$-) to non-passivating ($t$-time-dependent) SEI growth with increasing cycling currents.
These new insights extend our understanding of the influence of battery operation on battery life. This will aid in designing battery operation protocols for next-generation lithium-ion batteries.

Future works can extend our model for additional mechanical SEI deterioration, SEI heterogeneity, and lithium plating.
Implementing our model in two or three dimensions allows for an in-depth analysis of further mechanical SEI damaging like crack formation, spallation, or delamination. Moreover, this approach paves the way to better account for the heterogeneity and polycristallinity of the SEI.
As our model relies on lithium atoms as mediators for SEI growth, lithium plating, \textit{i.e.}, the accumulation of lithium on the anode, could be implemented in our model as additional degradation mode.
Furthermore, integrating our model in 3D full cell simulations, would capture the influence of heterogeneous electrodes on battery degradation. Especially graphite/silicon blend electrodes, which suffer from large mechanical differences, would profit from our degradation model. A solution to the accompanying computational challenges would be to simplify our SEI model to a 0D-version and implement it into a fully integrated electrochemo-mechanical FEM cell model \cite{Fang2019,Castelli2019,Castelli2021}.

  \section*{Acknowledgements}
    We gratefully acknowledge funding and support by the German Research Foundation (DFG) within the research training group SiMET under the project number 281041241/GRK2218. The support of the bwHPC initiative through the use of the JUSTUS HPC facility at Ulm University is acknowledged. This work contributes to the research performed at CELEST (Center for Electrochemical Energy Storage Ulm-Karlsruhe).

  \section*{Conflict of interest} 
    The authors declare no conflict of interest.

\bibliographystyle{unsrt}
\bibliography{literatur_abbr}

\end{document}